\documentclass
[pra,showpacs,a4paper,nofootinbib,superscriptaddress,twocolumn]{revtex4}%
\usepackage{amsmath}
\usepackage{amsfonts}
\usepackage{amssymb}
\usepackage{graphicx}
\usepackage{hyperref}%
\providecommand{\U}[1]{\protect\rule{.1in}{.1in}}
\pdfoutput=1

\begin{document}
\title{System-reservoir dynamics of quantum and classical correlations}
\author{J. Maziero}
\affiliation{Centro de Ci\^{e}ncias Naturais e Humanas, Universidade Federal do ABC, R.
Santa Ad\'{e}lia 166, 09210-170, Santo Andr\'{e}, S\~{a}o Paulo, Brazil}
\author{T. Werlang}
\affiliation{Departamento de F\'{\i}sica, Universidade Federal de S\~{a}o Carlos, Post
Office Box 676, 13565-905, S\~{a}o Carlos\textit{, }S\~{a}o Paulo\textit{, }Brazil}
\author{F. F. Fanchini}
\affiliation{Instituto de F\'{\i}sica Gleb Wataghin, Universidade Estadual de Campinas,
Post Office Box 6165, 13083-970, Campinas, S\~{a}o Paulo, Brazil}
\author{L. C. C\'{e}leri}
\affiliation{Centro de Ci\^{e}ncias Naturais e Humanas, Universidade Federal do ABC, R.
Santa Ad\'{e}lia 166, 09210-170, Santo Andr\'{e}, S\~{a}o Paulo, Brazil}
\author{R. M. Serra}
\affiliation{Centro de Ci\^{e}ncias Naturais e Humanas, Universidade Federal do ABC, R.
Santa Ad\'{e}lia 166, 09210-170, Santo Andr\'{e}, S\~{a}o Paulo, Brazil}

\begin{abstract}
We examine the system-reservoir dynamics of classical and quantum correlations
in the decoherence phenomenon within a two-qubit composite system interacting
with two independent environments. The most common noise channels (amplitude
damping, phase damping, bit flip, bit-phase flip, and phase flip) are
analyzed. By analytical and numerical analyses we find that, contrary to what
is usually stated in the literature, decoherence may occur without
entanglement between the system and the environment. We also show that, in
some cases, the bipartite quantum correlation initially present in the system
is completely evaporated and not transferred to the environments.

\end{abstract}

\pacs{03.65.Ta, 03.67.-a, 03.65.Yz}
\maketitle

\section{Introduction}

Until recently the quantum aspects of correlation were attributed to
inseparable quantum states \cite{Werner}, i.e., all nonclassical correlations
in a composite quantum state was regarded as entanglement. However, the
discovery that mixed separable (unentangled) states can also have nonclassical
correlation \cite{Nonlocal,OllZur} and that the use of such states can improve
performance in some computational tasks (compared to classical computing)
\cite{Caves,White} has opened a new perspective on the study and comprehension
of such correlations. The distinction between quantum and classical aspects of
correlation in a composite quantum state is an important issue in quantum
information theory (QIT). It is largely accepted that quantum mutual
information is a measure of the total correlation contained in a bipartite
quantum state \cite{GroPoWi,SchuWest}, but an outstanding question is how to
distinguish between the quantum and the classical aspects of the total
correlation. In view of the distinct nature of the correlations (quantum and
classical), it is reasonable to assume that they add in a simple way so that
quantum mutual information is the sum of the quantum and the classical
correlations \cite{HenVed,GroPoWi,Horo1,Horo2}.

To quantify the quantumness of the correlation contained in a bipartite
quantum state Olliver and Zurek \cite{OllZur} proposed a measure for quantum
correlation known as quantum discord and based on a distinction between QIT
and classical information theory (CIT). A related quantity concerning
classical correlation was proposed by Henderson and Vedral \cite{HenVed}. A
recent result that almost all quantum states have a nonvanishing quantum
discord \cite{Ferraro} shows up the relevance of studying such correlation.

At the core of the previous quantifiers of correlations is the one-partition
(one-side) measurement on a bipartite system. Thus, in a general case, those
quantifiers may be asymmetric with respect to the choice of subsystem to be
measured. A symmetrical quantifier of classical correlation based on
measurement over both partitions of a bipartite system was proposed in Ref.
\cite{DiVincenzo}. It was assumed that the classical correlation is quantified
by the maximum classical mutual information obtained by local measurements
over the two partitions of the system. An important point about these measures
of correlations is related to their computation. These measures are based on
extremization procedures over all possible measurements that can be performed
on the subsystems and thus constitute a difficult problem even numerically.
Actually, analytical solutions for the quantum discord were obtained recently
for a certain class of highly symmetrical states \cite{Lou, Sarandy, Maziero}.
Hence, an alternative, operational (without any extremization procedure)
quantifier is rather desirable.

One approach, also based on the disturbance that a measurement causes in the
system, was used in Ref. \cite{Lou1} where several quantifiers of correlations
were proposed. The author characterized classical states as those not
disturbed by a quantum measurement process. Another interesting proposal was
presented in Ref. \cite{Opp}. It was found that, while a certain quantity
related to the work that can be extracted from the environment using a
bipartite state is nonzero for all entangled states, it also need not vanish
for separable ones, being therefore a measure of quantum correlation.

Besides the characterization and quantification of classical and quantum
correlations, another important problem is the behavior of these correlations
under the action of decoherence. This phenomenon, mainly caused by the
injection of noise into the system and arising from its inevitable interaction
with the surrounding environment, is responsible for the loss of quantum
coherence initially present in the system. Recently it was noted
\cite{Werlang, Ferraro} that, for a certain class of states under Markovian
dynamics, the quantum discord only vanishes at asymptotic time, contrary to
what happens to entanglement, which can disappear at finite times \cite{Yu}.
These results show that the quantumness of correlation is more resistant to
the action of the environment than the entanglement itself. Although quantum
discord under decoherence does not exhibit sudden death, its dynamics may be
very peculiar, exhibiting sudden changes in behavior \cite{Maziero}.

Studying how decoherence affects the correlations in a two-qubit composite
system, Maziero and coworkers \cite{Maziero} recently proposed an operational
measure to quantify both classical and quantum correlations. This result rests
on the surprising fact that, for a suitable choice of the noise channel, the
classical correlation is not affected by the decoherence process, while the
quantum correlation is completely destroyed. Thus, the classical correlation
may be given by the quantum mutual information at asymptotic time
\cite{Maziero}.

In this article we are interested in the dynamics of system-reservoir
correlations under decoherence. We consider a noninteracting two-qubit system
under the influence of two independent environments. The most common noise
channels (amplitude damping, phase damping, bit flip, bit-phase flip, and
phase flip) are studied. By analytical and numerical analysis we find that,
contrary to what is usually stated in the literature, decoherence may occur
without entanglement between the system and the environment. We also show
that, in some cases, the bipartite quantum correlation initially present in
the system is completely evaporated and not transferred to the environments,
as can occur for entanglement under amplitude damping as reported in Ref.
\cite{Solano}.

The article is organized as follows. In Sec. II we discuss some proposed
measures of correlation: the mutual information in the realm of CIT and QIT,
the quantum discord and its generalization to a\ \textquotedblleft
two-side\textquotedblright\ measure of quantum correlation. We also present a
recently proposed operational measure (without any extremization procedure)
for both classical and quantum correlations. A brief review of the dynamics of
open quantum systems is presented in Sec. III and our results on the dynamics
of correlations under decoherence is presented in Sec. IV. We summarize our
conclusion and some possible avenues for future research in Sec. V.

\section{Measures of correlation}

\subsection{\textbf{Classical information theory}}

In CIT the mutual information measures the correlation between two random
variables $A$ and $B$ \cite{CIT}%
\begin{equation}
I_{c}(A\text{:}B)=\mathcal{H}(A)+\mathcal{H}(B)-\mathcal{H}(A,B),\label{CMI1}%
\end{equation}
where $\mathcal{H}(X)=-\sum\nolimits_{x}p_{x}\log p_{x}$ and $\mathcal{H}%
(A,B)=-\sum\nolimits_{a,b}p_{a,b}\log p_{a,b}$ are the Shannon entropies
(throughout this article all logarithms are taken to base $2$) for the
variable $X$ ($X=A,B$) and the joint system $AB$, respectively. $p_{a,b}$ is
the joint probability of the variables $A$ and $B$ assuming the values $a$ and
$b$, respectively, and $p_{a}=$ $\sum\nolimits_{b}p_{a,b}$\ $\left(
p_{b}=\sum\nolimits_{a}p_{a,b}\right)  $ is the marginal probability of the
variable $A$ ($B$) assuming the value $a$ ($b$).

From Bayes' rule \cite{CIT}, we can write the conditional probability%
\begin{equation}
p_{a|b}=\frac{p_{a,b}}{p_{b}},\label{cond}%
\end{equation}
and we also can express the joint entropy as $\mathcal{H}(A,B)=\mathcal{H}%
(A|B)+\mathcal{H}(B)$, where $\mathcal{H}(A|B)=-\sum_{a,b}p_{a,b}\log p_{a|b}$
is the conditional entropy of the variable $A$ given that variable $B$ is
known. Hence, the classical mutual information can also be expressed in terms
of the conditional entropy as%
\begin{equation}
J_{c}(A\text{:}B)=\mathcal{H}(A)-\mathcal{H}(A|B).\label{CMI2}%
\end{equation}
It is then straightforward to see that the two expressions (\ref{CMI1}) and
(\ref{CMI2}) for the classical mutual information are equivalent $\left[
I_{c}(A\text{:}B)-J_{c}(A\text{:}B)=0\right]  $.

\subsection{Quantum information theory}

In QIT, the extension of Eq. (\ref{CMI1}) to a bipartite quantum state
($\rho_{AB}$) is trivially obtained as \cite{NieChu, Bennetti, Vedral-Book}%
\begin{equation}
\mathcal{I}(\rho_{A:B})=S(\rho_{A})+S(\rho_{B})-S(\rho_{AB}), \label{QMI1}%
\end{equation}
where $S(\rho)=-\operatorname*{Tr}(\rho\log\rho)$ is the von Neumann entropy
and $\rho_{A(B)}=\operatorname*{Tr}_{B(A)}(\rho_{AB})$ is the reduced-density
operator of the partition $A$($B$). It is largely accepted that the quantum
mutual information $\mathcal{I}(\rho_{A:B})$ is the information-theoretic
measure of the total correlation (including both classical and quantum
correlations) in a bipartite quantum state \cite{GroPoWi, SchuWest}.

In the context of QIT, there is no quantum extension of Bayes' rule (for a
general state) \cite{Peres}. In fact, an analog of Bayes' rule can hold only
for composite quantum states without quantum correlation (a purely classically
correlated system). This departure from CIT arises from the nature of the
measurement process in quantum mechanics. Differently from the classical
scenario, the conditional probability Eq. (\ref{cond}) in QIT depends on which
observable is measured in the system $B$, since, in general, a quantum
measurement disturbs the system. This implies the nonequivalence \cite{OllZur}
of the quantum extensions of Eqs. (\ref{CMI1}) and (\ref{CMI2}).

\textit{One-side measures of correlation ---} To obtain a quantum version of
Eq. (\ref{CMI2}), let us consider a projective measurement $\Pi_{j}^{(B)}$ on
the subsystem $B$ of the composite state $\rho_{AB}$. The reduced state of
subsystem $A$ after the measurement, is given by
\[
\rho_{A}^{j}=\frac{1}{q_{j}}\operatorname*{Tr}\nolimits_{B}\left\{  \left(
\mathbf{1}_{A}\otimes\Pi_{j}^{(B)}\right)  \rho_{AB}\left(  \mathbf{1}%
_{A}\otimes\Pi_{j}^{(B)}\right)  \right\}  \text{,}%
\]
where $q_{j}=\operatorname*{Tr}_{AB}\left\{  \left(  \mathbf{1}_{A}\otimes
\Pi_{j}^{(B)}\right)  \rho_{AB}\right\}  $ is the probability for the
measurement of the $j$th state in subsystem $B$ and $\mathbf{1}_{A}$ is the
identity operator for subsystem $A$. For a complete set of projective
measurements $\left\{  \Pi_{j}^{(B)}\right\}  $, we can define the conditional
entropy of subsystem $A$, for a known subsystem $B$, as $S_{\left\{  \Pi
_{j}^{(B)}\right\}  }\left(  \rho_{A|B}\right)  \equiv\sum_{j}q_{j}S\left(
\rho_{A}^{j}\right)  $. So, we have the following quantum extension for Eq.
(\ref{CMI2})%
\begin{equation}
\mathcal{J}(\rho_{A:B})=S(\rho_{A})-S_{\left\{  \Pi_{j}^{(B)}\right\}
}\left(  \rho_{A|B}\right)  \text{.}\label{QMI2}%
\end{equation}

For a quantum correlated state, Eqs. (\ref{QMI1}) and (\ref{QMI2}) are not
equivalent. The difference
\begin{equation}
\mathcal{D}(\rho_{AB})\equiv\mathcal{I}(\rho_{A:B})-\max_{\left\{  \Pi
_{j}^{(B)}\right\}  }\mathcal{J}(\rho_{A\text{:}B})\label{Dis}%
\end{equation}
was called quantum discord by Olliver and Zurek \cite{OllZur}. One can say
that Eq. (\ref{Dis}) reveals the quantumness of the correlation between the
partitions $A$ and $B$ since it\ shows the departure between QIT and CIT. We
note that the nonclassical correlation captured by the quantum discord may be
present even in separable states \cite{OllZur}.

A quantum composite state may also have a classical correlation $\mathcal{C}%
(\rho_{AB})$, which for bipartite quantum states may be quantified via the
measure proposed by Henderson and Vedral \cite{HenVed}%
\begin{equation}
\mathcal{C}(\rho_{AB})\equiv\underset{\left\{  \Pi_{j}^{(B)}\right\}  }{\max
}\left[  S(\rho_{A})-S_{\left\{  \Pi_{j}^{(B)}\right\}  }(\rho_{\left.
A\right\vert B})\right]  ,\label{HV}%
\end{equation}
where the maximum is taken over the complete set of projective measurements
$\left\{  \Pi_{j}^{(B)}\right\}  $ on subsystem $B$. We consider only
projective measurements, rather than the more general positive operator-valued
measure (POVM) used in the original definition \cite{HenVed} of Eq.
(\ref{HV}). In fact, Hamieh and cowrokers \cite{Hamieh} showed that, for a
two-qubit system, the projective measurement is the POVM that maximizes Eq.
(\ref{HV}). For the purposes of this article we will only need to compute
correlations between two qubits, justifying the restricted set of measurements.

From the previous definitions, it follows immediately that $\mathcal{D}%
(\rho_{AB})+\mathcal{C}(\rho_{AB})=\mathcal{I}(\rho_{A:B})$, as expected. For
pure states, we have a special situation where the quantum discord is equal to
the entropy of entanglement and also equal to the Henderson-Vedral classical
correlation. In other words, $\mathcal{D}(\rho_{AB})=\mathcal{C}(\rho
_{AB})=\left.  \mathcal{I}(\rho_{A:B})\right/  2$ \cite{GroPoWi, HenVed}. In
this case, the total amount of quantum correlation is captured by an
entanglement measure. On the other hand, for mixed states, the entanglement is
only a part of this nonclassical correlation \cite{OllZur, White, Caves}.

It is worth mentioning that, for a general state, the quantum discord in Eq.
(\ref{Dis}) and also the (one-side) classical correlation of Eq. (\ref{HV})
may be asymmetric with respect to the choice of system to be measured. It can
be verified that, for states with maximally mixed marginals
($\operatorname*{Tr}_{A(B)}\rho_{AB}\propto\boldsymbol{1}_{B(A)}$),
$\mathcal{D}(\rho_{AB})$ and $\mathcal{C}(\rho_{AB})$ are symmetric under the
interchange $A\leftrightarrow B$.

\textit{Two-side measures of correlation --- }Besides \textquotedblleft
one-side\textquotedblright\ measures of quantum (\ref{Dis}) and classical
(\ref{HV}) correlations, we can define \textquotedblleft
two-side\textquotedblright\ measures for these correlations \cite{DiVincenzo,
Piani}. The classical correlation in a composite bipartite system can be
expressed as the \ \textquotedblleft maximum classical mutual
information\textquotedblright\ that can be obtained by local measurements on
both partitions of a composite state \cite{DiVincenzo}%
\begin{equation}
\mathcal{K}(\rho_{AB})\equiv\underset{\left\{  \Pi_{j}^{(A)}\otimes\Pi
_{j}^{(B)}\right\}  }{\max}\left[  I_{c}(\rho_{A:B})\right]  ,
\label{twosideCC}%
\end{equation}
where $I_{c}(\rho_{A:B})$ is the classical mutual information defined in Eq.
(\ref{CMI1}), $\mathcal{H}(A),$ $\mathcal{H}(B),$ $\mathcal{H}(A,B)$ being the
entropies of the probability distribution of the subsystems ($A$ and $B)$ and
the composite system ($AB$) resulting from a set of local projective
measurements $\Pi_{j}^{(A)}\otimes\Pi_{j}^{(B)}$ on both subsystems. Hence, a
two-side measure of quantum correlation can be defined as%
\begin{equation}
\mathcal{Q}(\rho_{AB})\equiv\mathcal{I}(\rho_{A:B})-\mathcal{K}(\rho_{AB}).
\label{twosideQC}%
\end{equation}

For composite states of two qubits with maximally mixed marginals, we have
numerically verified that the quantum discord (\ref{Dis}) is identical to the
two-side measure of quantum correlation (\ref{twosideQC}) [i.e. $\mathcal{D}%
(\rho_{AB})=\mathcal{Q}(\rho_{AB})$ and also $\mathcal{K}(\rho_{AB}%
)=\mathcal{C}(\rho_{AB})$].

\textit{Operational measures of correlation --- }Recently for a two-qubit
system, some of us proposed an operational measure of quantum and classical
correlations based on the dynamics of these correlations under decoherence
\cite{Maziero}. It was shown that, under suitable conditions, the classical
correlation is unaffected by decoherence. Such dynamic behavior leads to an
operational measure of both classical and quantum correlations that can be
computed without any extremization procedure. This can be done by sending the
component parts of a composed state through local channels that preserve its
classical correlation so that the quantum correlation $\mathbb{Q}(\rho_{AB})$
will be given simply by the difference between the state mutual information
$\mathcal{I}(\rho_{A:B})$ and the completely decohered mutual information
$\mathcal{I}\left[  \varepsilon(\rho_{A:B})\right]  $%
\begin{equation}
\mathbb{Q}(\rho_{AB})\equiv\mathcal{I}(\rho_{A:B})-\mathcal{I}\left[
\varepsilon(\rho_{A:B})\right]  ,\label{opQC}%
\end{equation}
since the classical correlation, $\mathbb{C}(\rho_{AB})$, present in
$\rho_{AB}$ is given by%
\begin{equation}
\mathbb{C}(\rho_{AB})=\mathcal{I}\left[  \varepsilon(\rho_{A:B})\right]
.\label{opCC}%
\end{equation}
Here, $\varepsilon(\rho_{AB})$ represents the evolved state of the system
under suitable local decoherence channels, described as a completely positive
trace-preserving map $\varepsilon\left(  \cdot\right)  $, in the asymptotic
time \cite{Maziero}. The choosing of suitable channels that preserve the
classical correlation is the challenging part of this measure. Until now, this
problem was solved only for a given class of composite states of two qubits
with maximally mixed marginals \cite{Maziero}.

\section{Dynamics of open quantum systems}

Let us briefly review the theory of open quantum systems (for a complete
treatment see Ref.  \cite{BOpen}). The time evolution of a general closed
quantum system is governed by the Liouville-von Neumann equation (we will use
natural units, such that $\hslash=1$)%
\begin{equation}
\dot{\rho}(t)=-i\left[  H,\rho(t)\right]  ,\label{D1}%
\end{equation}
where $\rho$ and $H$ are the density operator and the Hamiltonian of the
system, respectively. This equation implies that the evolution is unitary.
However, in a realistic scenario the system of interest ($S$) --- hereafter
referred to as just system --- interacts with its surrounding environment
($E$) (also referred to as reservoir). To account for this unavoidable
interaction, which is often the major source of noise introduced into the
system, we can rewrite the complete Hamiltonian as%
\[
H=H_{S}+H_{E}+H_{I},
\]
where $H_{S}$ and $H_{E}$ are the bare system and environment Hamiltonians,
respectively, and $H_{I}$ the interaction Hamiltonian. While it is true that
the whole system ($S+E$) still respects Eq. (\ref{D1}) (the density operator
$\rho=\rho_{SE}$ now also includes the variables of the environment), in
general, we are only interested in obtaining an effective dynamic equation for
the $S$ variables. This may be done by taking the partial trace of Eq.
(\ref{D1}) over the $E$ variables. Then the reduced system dynamics is
governed by%
\begin{equation}
\dot{\rho}_{S}(t)=-i\operatorname*{Tr}\nolimits_{E}\left\{  \left[
H,\rho_{SE}(t)\right]  \right\}  ,\label{D2}%
\end{equation}
where $\rho_{S}=\operatorname*{Tr}\nolimits_{E}\left(  \rho_{SE}\right)  $ is
the reduced-density operator of the system. This evolution is not, in general,
unitary and leads to the phenomenon known as decoherence \cite{Max}. If we
assume that the environment is Markovian (which implies a large number of
degrees of freedom) and initially uncorrelated with the system $S$ ($\rho
_{SE}(0)=\rho_{S}(0)\otimes\rho_{E}(0)$, $\rho_{E}$ being the reduced-density
operator of the environment), Eq. (\ref{D2}) can be written as a sum of
operators acting only on the system%
\begin{align*}
\dot{\rho}_{S}(t) &  =-i\left[  H_{S},\rho_{S}(t)\right]  \\
&  -\sum\limits_{i,j}\gamma_{i,j}\left[  \rho_{S}(t)L_{i}L_{j}+L_{i}L_{j}%
\rho_{S}(t)\right.  \\
&  \left.  -L_{j}\rho_{S}(t)L_{i}\right]  +H.c.,
\end{align*}
where $L_{j}$ is the so-called Lindblad operator and $\gamma_{i,j}$ is a
constant that depends on the specific decoherence process. This is the well
known master equation approach for open quantum systems \cite{BOpen}. It is
important to note that this approach depends on the perturbation theory for
the system-environment coupling parameter, which implies that it is valid only
in the weak coupling regime, i.e., when $S$ is nearly closed.

Although the master equation approach is widely used, specially in quantum
optics \cite{Zoller}, there is another way to treat open quantum systems,
which is more appropriate for our purposes. We will only sketch this approach
in what follows (a complete treatment can be found in Ref. \cite{NieChu}). The
formal solution of Eq. (\ref{D1}) can be written in the form%
\begin{equation}
\rho_{SE}(t)=\mathbf{U}(t)\rho_{SE}(0)\mathbf{U}^{\dag}(t)\text{,}\label{D3}%
\end{equation}
where $\mathbf{U}(t)$ is the unitary evolution operator generated by the total
($S+E$) Hamiltonian. The partial trace over the environment variables defines
a completely positive map $\varepsilon\left(  \cdot\right)  $\ for all
classically correlated system-environment initial states \cite{Rosario} that
describes the evolution of the system $S$ under the action of the environment
$E$
\begin{equation}
\varepsilon\left(  \rho_{S}\right)  =\operatorname*{Tr}\nolimits_{E}\left\{
\mathbf{U}(t)\rho_{SE}(0)\mathbf{U}^{\dag}(t)\right\}  \text{.}\label{D4}%
\end{equation}
The map $\varepsilon$ is a quantum operation, not necessarily unitary, mapping
density operators into density operators and this is the reason that
$\varepsilon$ is a completely positive map\footnote{If $\Lambda_{AB}$ is a
positive map and $\rho_{AB}$ is the density operator of the composite system
$AB$, then $\tilde{\rho}_{AB}=\Lambda_{AB}\left(  \rho_{AB}\right)  $ is also
a valid density operator (all its eigenvalues are non-negative). If
$\Lambda_{B}$ is a completely positive map, $\tilde{\rho}_{AB}=I_{A}%
\otimes\Lambda_{B}\left(  \rho_{AB}\right)  $ is also a valid density
operator.}. Assuming that the system and the environment are initially
uncorrelated $\left[  \rho_{SE}(0)=\rho_{S}(0)\otimes\rho_{E}(0)\right]  $, we
can rewrite Eq. (\ref{D4}) in the so-called operator-sum representation%
\begin{equation}
\varepsilon\left(  \rho_{S}\right)  =\sum_{k}\Gamma_{k}\rho_{S}\Gamma
_{k}^{\dag}\text{,}\label{D5}%
\end{equation}
with the Kraus operators $\Gamma_{k}(t)=\left.  _{E}\left\langle k\right\vert
\right.  \mathbf{U}(t)\rho_{E}|k\rangle_{E}$ acting only on the state space of
system $S$, $\left\{  |k\rangle_{E}\right\}  $ being an orthonormal basis for
the environment. The Kraus operators satisfy the completeness relation
$\sum_{k}\Gamma_{k}^{\dag}\Gamma_{k}=\boldsymbol{1}$, yielding a map
$\varepsilon$ that is trace-preserving\footnote{In fact, this condition can be
generalized to include non-trace-preserving maps, such as a measurement
process. In this case, we have $\sum_{k}\Gamma_{k}^{\dag}\Gamma_{k}%
\leq\boldsymbol{1}$ \cite{NieChu}.}. This definition of the Kraus operators is
not unique. If we adopt a different basis to compute the trace in Eq.
(\ref{D4}), we obtain a different set of \textit{equivalent} operators in the
sense that both sets generate the same dynamics for the system (the same
operation). This can be seen from the fact that these two sets of operators
are connected to each other by a unitary transformation. Moreover, it can be
shown that, assuming a Markovian environment, Eq. (\ref{D5}) leads to the same
master equation as that obtained from Eq. (\ref{D2}). From these
considerations we can see that the operator-sum representation is more general
than the master equation approach in the sense that the former can be applied
even if the environment has only a few degrees of freedom. Another advantage
of this tool is that it can be applied, in a simultaneous way, to a large
range of physical systems, since Eq. (\ref{D5}) does not include specific
details of the environment, providing us with a quite general dynamic equation
for the system $S$.

To generalize this formalism for the case in which the system $S$ is composed
of more than one part, we must specify which type of environment we are
dealing with. Let us consider two types of environment (i) global and (ii)
local. In case $(i)$, the interaction of all parts of $S$ with the same
environment may lead, in principle, to an increase in correlations between the
parts of the system due to \textquotedblleft nonlocal
interactions\textquotedblright\ mediated by the environment \cite{Global}. In
case (ii), each part of $S$ interacts with its local, independent environment.
It is clear that correlations cannot then be increased between the parts of
the system by interaction with the environment. For the case (ii), regarding
$N$ system parts and $N$ independent environments, Eq. (\ref{D5}) immediately
becomes%
\begin{equation}
\varepsilon\left(  \rho_{S}\right)  =\sum_{k_{1},...,k_{N}}\Gamma_{k_{1}%
}^{(1)}\otimes\cdot\cdot\cdot\otimes\Gamma_{k_{N}}^{(N)}\rho_{S}\Gamma_{k_{1}%
}^{(1)\dag}\otimes\cdot\cdot\cdot\otimes\Gamma_{k_{N}}^{(N)\dag}%
\text{.}\label{D6}%
\end{equation}
Here $\Gamma_{k_{\alpha}}^{(\alpha)}$ is the $k_{\alpha}$th Kraus operator for
the environment acting on subsystem $\alpha$. This can be verified directly
from the fact that the total evolution operator in Eq. (\ref{D4}) can be
written in the product form $\mathbf{U}(t)=\mathbf{U}_{1}(t)\otimes
\mathbf{U}_{2}(t)\otimes\cdot\cdot\cdot\otimes\mathbf{U}_{N}(t)$.

The decoherence process can also be represented by a map in terms of the
complete system-environment state. Let $\left\{  |\zeta_{l}\rangle
_{S}\right\}  $, with $l=1,\cdots,d$, be a complete basis for $S$. Then, there
are, at most, $d^{2}$ Kraus operators \cite{Leung03} and the dynamics of the
complete system can be represented by the following map \cite{Salles08}
\begin{widetext}
\begin{align}
|\zeta_{1}\rangle_{S}\otimes|0\rangle_{E} &  \rightarrow\Gamma_{0}|\zeta
_{1}\rangle_{S}\otimes|0\rangle_{E}+\cdots+\Gamma_{d^{2}-1}|\zeta_{1}%
\rangle_{S}\otimes|d^{2}-1\rangle_{E}\nonumber\\
|\zeta_{2}\rangle_{S}\otimes|0\rangle_{E} &  \rightarrow\Gamma_{0}|\zeta
_{2}\rangle_{S}\otimes|0\rangle_{E}+\cdots+\Gamma_{d^{2}-1}|\zeta_{2}%
\rangle_{S}\otimes|d^{2}-1\rangle_{E}\nonumber\\
&  \vdots\nonumber\\
|\zeta_{d}\rangle_{S}\otimes|0\rangle_{E} &  \rightarrow\Gamma_{0}|\zeta
_{d}\rangle_{S}\otimes|0\rangle_{E}+\cdots+\Gamma_{d^{2}-1}|\zeta_{d}%
\rangle_{S}\otimes|d^{2}-1\rangle_{E}\text{,}\label{MapGen}%
\end{align}
\end{widetext}given that%
\[
\mathbf{U}_{SE}|\zeta_{l}\rangle_{S}\otimes|0\rangle_{E}=\sum_{k}\Gamma
_{k}|\zeta_{l}\rangle_{S}\otimes|k\rangle_{E}\text{.}%
\]

Here we will be interested only in the case of local, independent environments.

\section{Correlation Dynamics under Decoherence}

In this section we will investigate the correlation dynamics of a two-qubit
system $\rho_{AB}$ under the action of two local environments. The most
general two-qubit state can be written in the form \cite{Jaeger}%
\begin{equation}
\rho_{AB}\left(  0\right)  =\frac{1}{4}\sum_{i,j=0}^{3}c_{i,j}\sigma_{i}%
^{(A)}\otimes\sigma_{j}^{(B)},\label{state}%
\end{equation}
where $\sigma_{i}^{(k)}$ is the standard Pauli matrix in direction $i$
($i=1,2,3$) acting on the space of subsystem $k$ ($k=A,B$), $\sigma_{0}%
^{(k)}=\boldsymbol{1}_{k}$ being the identity operator for the partition $k$
and $c_{i,j}$ are real coefficients that satisfy both positivity and
normalization of $\rho_{AB}$. Our goal here is to study the dynamics of
classical and quantum correlations as well as the entanglement within the
possible bipartitions of the complete system (the system of interest plus the
environments) under the action of several noise channels. We consider the most
common decoherence channels (i.e., amplitude damping, phase damping, bit flip,
bit-phase flip, and phase flip).

In what follows, we will consider a system $S$ constituted by the two qubits
$A$ and $B$, each of them interacting independently with its own environment
$E_{A}$ and $E_{B}$, respectively.

\subsection{Amplitude-damping}

The amplitude-damping channel is a\ classical noise process describing the
dissipative interaction between the system and the environment. There is an
exchange of energy between $S$\textbf{ }and\textbf{ }$E$, such that $S$ is
driven into thermal equilibrium with $E$. This channel may be modeled by
treating $E$ as a large collection of independent harmonic oscillators
interacting weakly with $S$, as in the case of the spontaneous emission of an
excited atom in the vacuum electromagnetic field (the reservoirs are at zero
temperature, i.e., in the vacuum state) \cite{BOpen, Max, Zoller}.

The action of a dissipative channel over one qubit can be represented by the
following phenomenological map (from Eq. \ref{MapGen})%
\begin{subequations}
\begin{align}
|0\rangle_{S}|0\rangle_{E} &  \rightarrow|0\rangle_{S}|0\rangle_{E}%
\label{Mapl1}\\
|1\rangle_{S}|0\rangle_{E} &  \rightarrow\sqrt{q}|1\rangle_{S}|0\rangle
_{E}+\sqrt{p}|0\rangle_{S}|1\rangle_{E},\label{Mapl2}%
\end{align}
where $|0\rangle_{S}$ is the ground and $|1\rangle_{S}$ the excited qubit
state. $|0\rangle_{E}$ and $|1\rangle_{E}$ describe the states of the
environment with no excitation and one excitation distributed over all its
modes, respectively. Equation (\ref{Mapl1}) describes the fact that if the
system and the environment start in the ground state, there is no dynamic
evolution. Equation (\ref{Mapl2}) tells us that if the qubit starts in the
excited state, there is a probability $q\equiv1-p$ that it will remain in this
state and a probability $p$ that it will decay. We are using $p$ to describe
these probabilities as a parametrization of time such that $p\in\left[
0,1\right]  $. The exact dependence of $p$ on time will depend, of course, on
the specific model for the environment as well as on the system under
consideration. For example, if we consider an infinite bosonic environment
interacting with a two-level fermionic system under the Markovian
approximation, $p$ will be a decreasing exponential function of time. On the
other hand, if we are dealing with an (non-Markovian) environment with a small
number of degrees of freedom, $p$ may be an oscillatory function of time. The
advantage of using $p$ instead of an explicit function of time is the
possibility of describing a wide range of physical systems in the same dynamics.

We can get a geometrical picture of the action of this channel by considering
the Bloch sphere representation of one qubit interacting with an infinite
bosonic reservoir at zero temperature (a tractable model for the
amplitude-damping channel). As already noted, in this case $p$ is an
exponentially decreasing function of time and the action of the channel is
then to move every point of the unit sphere to the pole where the
$|0\rangle_{S}$ state is located. In other words, in the asymptotic limit the
whole sphere is compressed to a single point located at the (lower energy) pole.

From Eqs. (\ref{Mapl1}) and (\ref{Mapl2}), we can see that the Kraus operators
describing the amplitude-damping channel are given \cite{NieChu} by%
\end{subequations}
\begin{equation}
\Gamma_{0}^{(k)}=%
\begin{bmatrix}
1 & 0\\
0 & \sqrt{q}%
\end{bmatrix}
\text{, \ }\Gamma_{1}^{(k)}=%
\begin{bmatrix}
0 & \sqrt{p}\\
0 & 0
\end{bmatrix}
\text{,}\label{KAD}%
\end{equation}
where $k=A,B$ labels the two distinct environments (one for each qubit).

Let us start by studying the correlations between the various bipartitions of
the whole system ($S+E$), assuming that the initial total density operator is
given by%
\begin{equation}
\rho_{ABE_{A}E_{B}}=\frac{1}{4}\left(  \sum_{i=0}^{3}c_{i}\sigma_{i}%
^{(A)}\otimes\sigma_{i}^{(B)}\right)  \otimes|00\rangle_{E_{A}E_{B}},
\label{init}%
\end{equation}
where $|00\rangle_{E_{A}E_{B}}$ is the vacuum (zero temperature) state of the
environments $E_{A}$ and $E_{B}$, in which the qubits $A$ and $B$,
respectively, are immersed. The coefficients $c_{i}$ ($c_{0}\equiv1$), are
real constants constrained in such a way that $\rho_{ABE_{A}E_{B}}$ is
positive and normalized. We note that the state of the system $AB$ in Eq.
(\ref{init}) represents a considerable class of states including the Werner
($\left\vert c_{1}\right\vert =\left\vert c_{2}\right\vert =\left\vert
c_{3}\right\vert =\alpha$) and Bell basis states ($\left\vert c_{1}\right\vert
=\left\vert c_{2}\right\vert =\left\vert c_{3}\right\vert =1$).

Although we can compute from Eqs. (\ref{Mapl1}), (\ref{Mapl2}) and
(\ref{init}), the total density operator, $\rho_{ABE_{A}E_{B}}\left(
p\right)  $ evolved under the action of the amplitude-damping channel in a
straightforward way, it is too cumbersome to be shown here. As we are
interested in the correlations between the parts in the various bipartitions
of the complete system, we will only need the corresponding reduced matrices.
The reduced-density operator for the partition $AB$, obtained by taking the
partial trace of $\rho_{ABE_{A}E_{B}}\left(  p\right)  $ over the degrees of
freedom of the reservoir $\rho_{AB}\left(  p\right)  =\operatorname*{Tr}%
\nolimits_{E_{A}E_{B}}[\rho_{ABE_{A}E_{B}}\left(  p\right)  ]$, in the
computational basis $\left\{  \left\vert 00\right\rangle _{kl},\left\vert
01\right\rangle _{kl},\left\vert 10\right\rangle _{kl},\left\vert
11\right\rangle _{kl}\right\}  $ for the partition $kl$ ($k=A$, $l=B$), is
given by \begin{widetext}
\begin{equation}
\rho_{AB}\left(  p\right)  =\frac{1}{4}%
\begin{bmatrix}
\left(1+p\right)^{2} + \left(1-p\right)^{2}c_{3} & 0 & 0 & q\left(  c_{1}-c_{2}\right) \\
0 & \left( 1-c_{3}\right)q +\left( 1+c_{3}\right)pq & q\left(  c_{1}+c_{2}\right)  & 0\\
0 & q\left(  c_{1}+c_{2}\right)  & \left( 1-c_{3}\right)q +\left( 1+c_{3}\right)pq & 0\\
q\left(  c_{1}-c_{2}\right)  & 0 & 0 & q^{2}\left(  1+c_{3}\right)
\end{bmatrix}
. \label{r1ad}%
\end{equation}
\end{widetext}

For the partitions $AE_{A}$ and $AE_{B}$, the reduced-density operators read%
\begin{equation}
\rho_{AE_{A}}(p)=\frac{1}{2}%
\begin{bmatrix}
1 & 0 & 0 & 0\\
0 & p & \sqrt{pq} & 0\\
0 & \sqrt{pq} & q & 0\\
0 & 0 & 0 & 0
\end{bmatrix}
\label{r2ad}%
\end{equation}
and \begin{widetext}
\begin{equation}
\rho_{AE_{B}}\left(  p\right)  =\frac{1}{4}%
\begin{bmatrix}
\left(  1+c_{3}\right)  \left(  1+pq\right)  +1-c_{3} & 0 & 0 & \left(
c_{1}-c_{2}\right)  \sqrt{pq}\\
0 & \left(  1-c_{3}\right)  p+\left(  1+c_{3}\right)  p^{2} & 0 & 0\\
0 & 0 & \left(  1-c_{3}\right)  q+\left(  1+c_{3}\right)  q^{2} & 0\\
\left(  c_{1}-c_{2}\right)  \sqrt{pq} & 0 & 0 & \left(  1+c_{3}\right)  pq
\end{bmatrix}
,\label{r3ad}%
\end{equation}
\end{widetext}respectively.

Finally, for the partition $E_{A}E_{B}$, obtained by tracing out the system
degrees of freedom, we obtain \begin{widetext}
\begin{equation}
\rho_{E_{A}E_{B}}\left(  p\right)  =\frac{1}{4}%
\begin{bmatrix}
\left(1+q\right)^{2}+\left(1-q\right)^{2}c_{3} & 0 & 0 & \left(  c_{1}-c_{2}\right)  p\\
0 & \left(  1-c_{3}\right)p + \left(  1+c_{3}\right)pq & \left(  c_{1}+c_{2}\right)  p & 0\\
0 & \left(  c_{1}+c_{2}\right)  p & \left(  1-c_{3}\right)p + \left(  1+c_{3}\right)pq & 0\\
\left(  c_{1}-c_{2}\right)  p & 0 & 0 & \left(  1+c_{3}\right)  p^{2}%
\end{bmatrix}
. \label{r4ad}%
\end{equation}
\end{widetext}Due to the inherent symmetry of the system, the density matrix
representing the partition $BE_{B}$ is identical to that for the partition
$AE_{A}$, thus leading to the same dynamics. The same symmetry is exhibited
between the partitions $AE_{B}$ and $BE_{A}$.

Due to the $X$ structure of the density matrices (\ref{r1ad}) through
(\ref{r4ad}), there is a simple closed expression for the concurrence present
in all bipartitions%
\begin{equation}
C(p)=2\max\left\{  0,\Lambda_{1}(p),\Lambda_{2}(p)\right\}  ,\label{Conc1}%
\end{equation}
with $\Lambda_{1}(p)=\left\vert \rho_{14}\right\vert -\sqrt{\rho_{22}\rho
_{33}}$ and $\Lambda_{2}(p)=\left\vert \rho_{23}\right\vert -\sqrt{\rho
_{11}\rho_{44}}$. For the classical Eq. (\ref{twosideCC}) and quantum Eq.
(\ref{twosideQC})\ correlations we have no analytical expression; however,
numerical analysis is possible. To this end, we will consider a Werner initial
state, where $c_{1}=c_{2}=c_{3}=-\alpha$ ($0\leq\alpha\leq1$). In Fig. 1, we
show the dynamics of the correlations for the partition $AB$. First we note
that both classical ($\mathcal{K}$) and quantum ($\mathcal{Q}$) correlations
only vanish in the asymptotic limit ($p=1$), while the entanglement suffers a
sudden death (SD) at a certain parametrized time $p_{SD}$ \cite{Yu,
Davidovich}. This can be seen directly from Eq. (\ref{Conc1}). On the other
hand, such a system exhibits a sudden birth (SB) of entanglement between the
reservoirs ($E_{A}E_{B}$) \cite{Solano}.%

\begin{figure}
[h]
\begin{center}
\includegraphics[
natheight=8.645500in,
natwidth=10.562800in,
height=7.4641cm,
width=8.9227cm
]%
{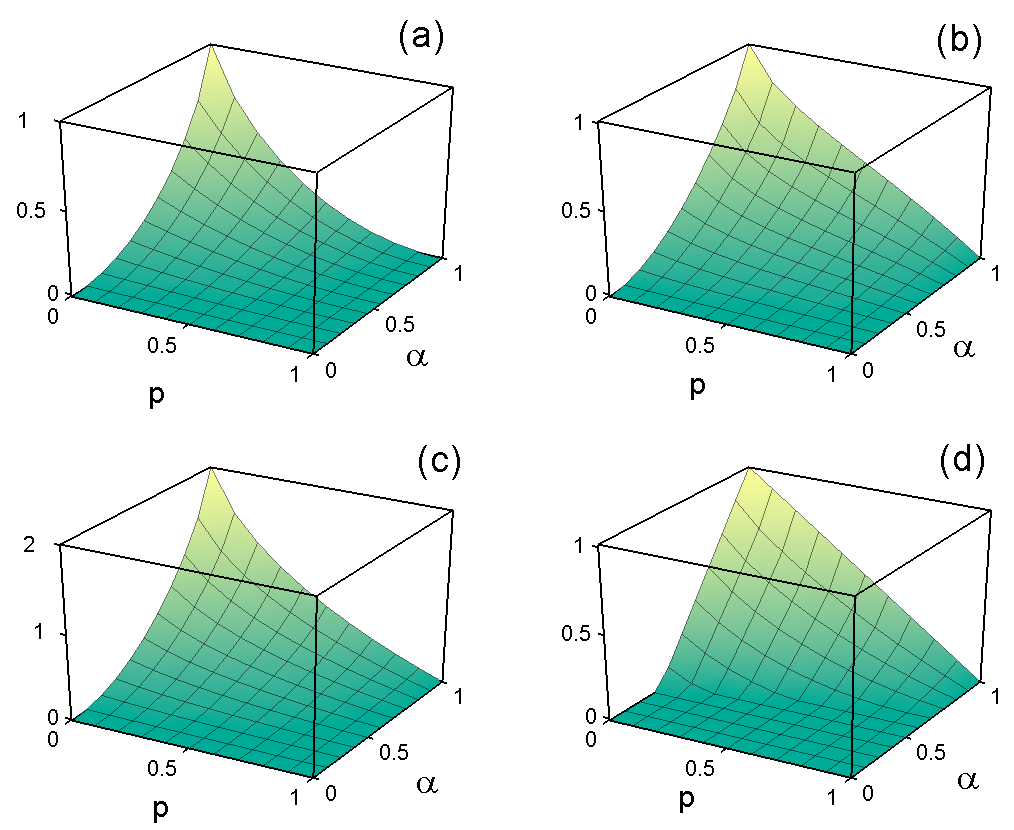}%
\caption{(Color online) Correlation dynamics for the amplitude-damping
channel, considering partition $AB$ for the Werner initial state. (a)
Classical correlation [Eq. (\ref{twosideCC})], (b) quantum correlation [Eq.
(\ref{twosideQC})], (c) mutual information [Eq. (\ref{QMI1})], and (d)
concurrence [Eq. (\ref{Conc1})]. }%
\end{center}
\end{figure}

The fact that the \textquotedblleft entanglement sudden
death\textquotedblright\ between the qubits and \textquotedblleft entanglement
sudden birth\textquotedblright\ between the reservoirs may occur at different
instants was first reported in Ref. \cite{Solano} and is shown in Figs. 2 and
3. On the other hand, as we can see in\ Figs. 1 and 2, contrary to what
happens to entanglement, the vanishing of the classical and quantum
correlations between the parts of system $AB$ is accompanied, simultaneously,
by the creation of such correlations between the reservoirs. Moreover, in Fig.
1 we see that, although the entanglement in partition $AB$ disappears at a
finite time, the quantum correlation $\mathcal{Q}$ vanishes only
asymptotically, as previously noted in Ref. \cite{Werlang}.%

\begin{figure}
[h]
\begin{center}
\includegraphics[
natheight=8.416400in,
natwidth=10.625100in,
height=7.2269cm,
width=8.9754cm
]%
{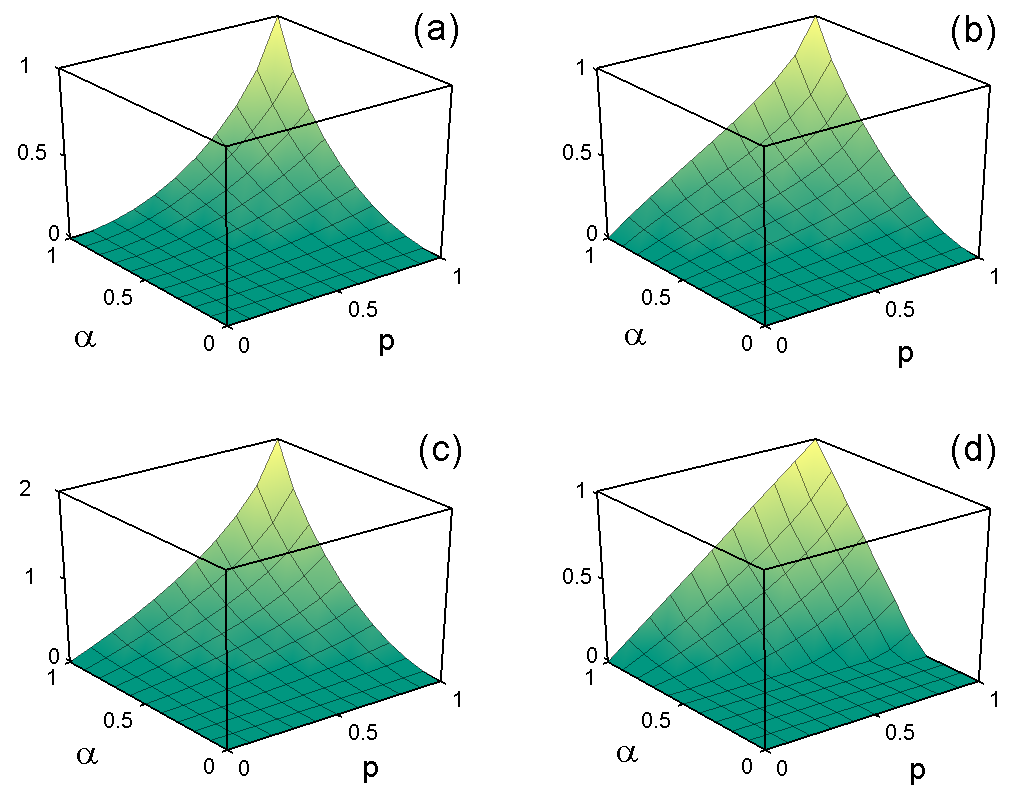}%
\caption{(Color online) Correlation dynamics for the amplitude-damping channe,
considering partition $E_{A}E_{B}$ for the Werner initial state. (a) Classical
correlation [Eq. (\ref{twosideCC})], (b) quantum correlation [Eq.
(\ref{twosideQC})], (c) mutual information [Eq. (\ref{QMI1})], and (d)
concurrence [Eq. (\ref{Conc1})].}%
\end{center}
\end{figure}
\begin{figure}
[h]
\begin{center}
\includegraphics[
natheight=4.599900in,
natwidth=11.400000in,
height=3.7145cm,
width=9.1028cm
]%
{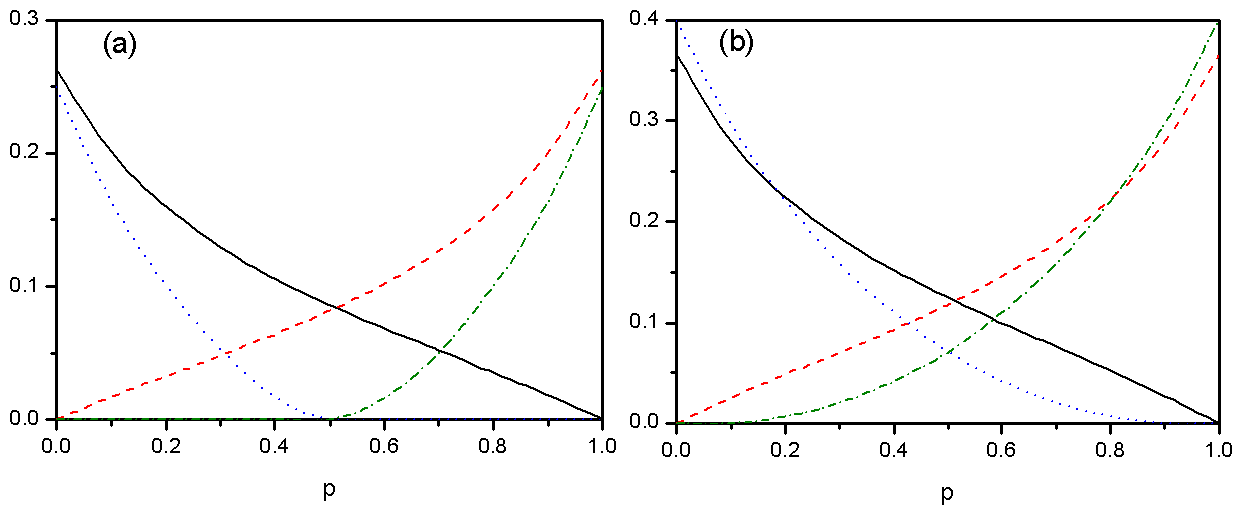}%
\caption{(Color online) Correlation dynamics for the two-qubit system with the
amplitude-damping channel for the Werner state with (a) $\alpha=0.5$ and (b)
$\alpha=0.6$. Quantum correlations given by Eq. (\ref{twosideQC}) for
partitions $AB$ (solid line) and $E_{A}E_{B}$ (dashed line) and entanglement
given by Eq. (\ref{Conc1}) for partitions $AB$ (dotted line) and $E_{A}E_{B}$
(dot-dashed line).}%
\end{center}
\end{figure}

In Figs. 4 and 5 we plot the dynamics of correlations for the partitions
$AE_{A}$ and $AE_{B}$, respectively. From these figures and from the fact that
$\rho_{AE_{A}}\left(  p\right)  =\rho_{BE_{B}}\left(  p\right)  $ and
$\rho_{AE_{B}}\left(  p\right)  =\rho_{BE_{A}}\left(  p\right)  $, we see that
each qubit has nonclassical correlation only with its own reservoir for all
values of $\alpha$ and $p$. Also, in the asymptotic limit, as expected, all
the correlations between the system and the reservoirs vanish due to the fact
that we considered the reservoir initially in the vacuum state.

%

\begin{figure}
[h]
\begin{center}
\includegraphics[
natheight=11.813300in,
natwidth=13.920000in,
height=5.7398cm,
width=8.0946cm
]%
{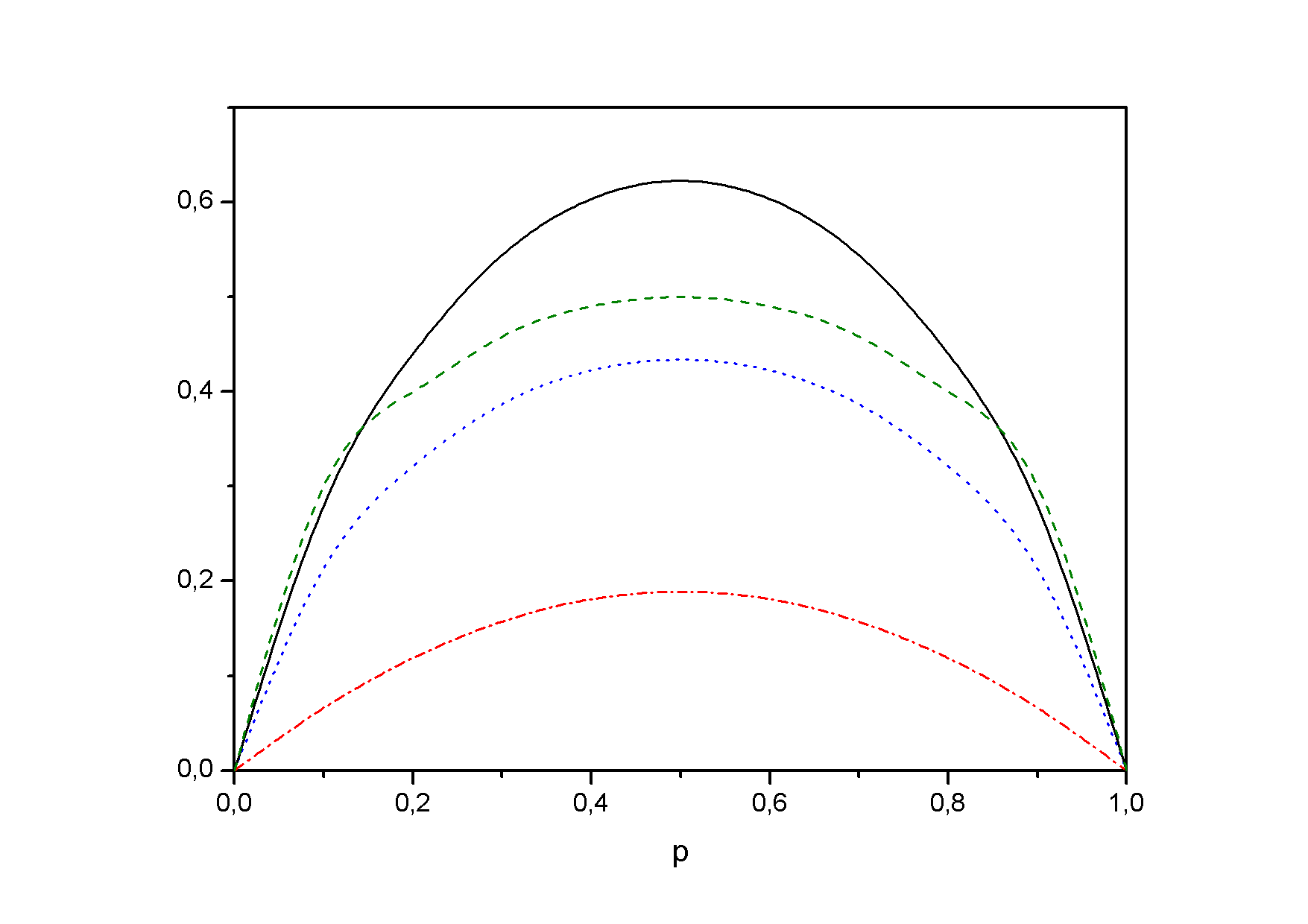}%
\caption{(Color online) Correlation dynamics for the amplitude-damping
channel, considering the partition $AE_{A}$ for the general state
(\ref{init}). Classical correlation (dot-dashed line) given by Eq.
(\ref{twosideCC}), quantum correlation (dotted line) given by Eq.
(\ref{twosideQC}), mutual information (solid line) given by Eq. (\ref{QMI1}),
and concurrence (dashed line) given by Eq. (\ref{Conc1}).}%
\end{center}
\end{figure}
%

\begin{figure}
[h]
\begin{center}
\includegraphics[
natheight=8.687900in,
natwidth=10.728900in,
height=7.3851cm,
width=9.1028cm
]%
{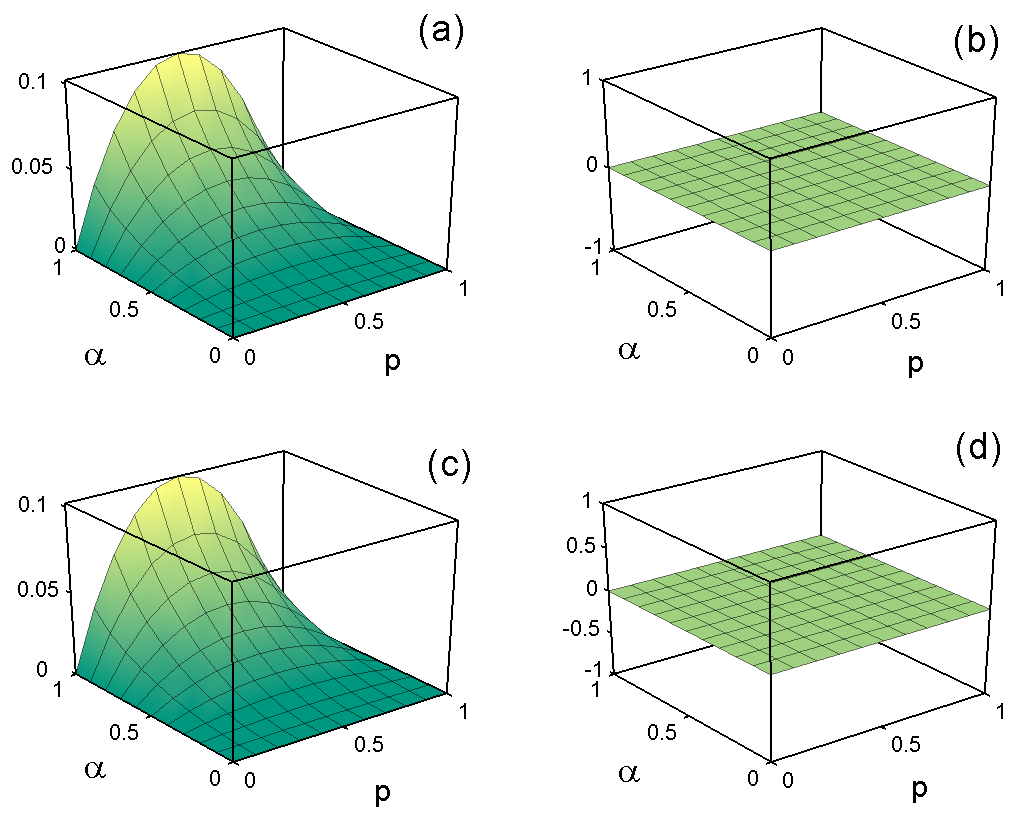}%
\caption{(Color online) Correlation dynamics for the amplitude-damping channel
considering the partition $AE_{B}$ for the Werner initial state. (a) Classical
correlation [Eq. (\ref{twosideCC})], (b) quantum correlation [Eq.
(\ref{twosideQC})], (c) mutual information [Eq. (\ref{QMI1})], and (d)
entanglement [Eq. (\ref{Conc1})]. }%
\end{center}
\end{figure}

\subsection{Phase-damping}

The phase-damping channel describes the loss of quantum coherence without loss
of energy. It leads to decoherence without relaxation. An example of a
physical system described by this channel is the random scattering of a photon
in a waveguide \cite{NieChu}. The map that describes the action of this
channel on a one-qubit system is given by%
\begin{align}
|0\rangle_{S}|0\rangle_{E} &  \rightarrow|0\rangle_{S}|0\rangle_{E}\nonumber\\
|1\rangle_{S}|0\rangle_{E} &  \rightarrow\sqrt{q}|1\rangle_{S}|0\rangle
_{E}+\sqrt{p}|1\rangle_{S}|1\rangle_{E}\text{.}\label{Map2}%
\end{align}
So that there is no exchange of energy between the system and the reservoir,
only the phase relations between the energy eigenstates of the system are lost
during time evolution. The Kraus operators describing the phase-damping
channel for the qubit $k$ ($k=A,B$) may be written as%
\[
\Gamma_{0}^{(k)}=%
\begin{bmatrix}
1 & 0\\
0 & \sqrt{q}%
\end{bmatrix}
\text{, \ }\Gamma_{1}^{(k)}=%
\begin{bmatrix}
0 & 0\\
0 & \sqrt{p}%
\end{bmatrix}
\text{.}%
\]
Assuming the initial state (\ref{init}), the evolved density operator of the
partition $AB$, obtained by tracing out the degrees of freedom of the
reservoirs, is given by%
\begin{equation}
\varepsilon\left(  \rho_{AB}\right)  =\frac{1}{4}%
\begin{bmatrix}
1+c_{3} & 0 & 0 & c^{-}q\\
0 & 1-c_{3} & c^{+}q & 0\\
0 & c^{+}q & 1-c_{3} & 0\\
c^{-}q & 0 & 0 & 1+c_{3}%
\end{bmatrix}
,\label{r1d}%
\end{equation}
where we defined $c^{\pm}=c_{1}\pm c_{2}$. The classical and quantum
correlations present in the reduced state (\ref{r1d}) can be computed
analytically through the measures (\ref{opCC}) and (\ref{opQC}) and are given
by \cite{Maziero}%

\begin{subequations}
\begin{align}
\mathcal{C}\left[  \varepsilon\left(  \rho_{AB}\right)  \right]   &
=\sum_{k=1}^{2}\frac{1+(-1)^{k}\chi}{2}\log_{2}\left[  1+(-1)^{k}\chi\right]
,\label{CC}\\
\mathcal{D}\left[  \varepsilon\left(  \rho_{AB}\right)  \right]   &
=2+\sum_{k=1}^{4}\lambda_{k}\log_{2}\lambda_{k}-\mathcal{C}\left[
\varepsilon(\rho_{AB})\right]  ,\label{CQ}%
\end{align}
where $\chi=\max\left\{  q^{2}\left\vert c_{1}\right\vert ,q^{2}\left\vert
c_{2}\right\vert ,\left\vert c_{3}\right\vert \right\}  $ and $\lambda_{k}$ is
the $k$th eigenvalue of the reduced-density matrix $\rho_{AB}\left(  p\right)
$ \cite{Maziero}. We can verify that $\mathcal{C}\left[  \varepsilon\left(
\rho_{AB}\right)  \right]  $ and $\mathcal{D}\left[  \varepsilon\left(
\rho_{AB}\right)  \right]  $ are symmetric under the interchange
$A\leftrightarrow B$. It was also numerically verified that the
\textquotedblleft one-side\textquotedblright\ measures of correlations
(\ref{HV}) and (\ref{Dis}) lead in this special case to the same values as the
\textquotedblleft two-side\textquotedblright\ measures (\ref{twosideCC}) and
(\ref{twosideQC}), respectively. Therefore, for this state, the quantum
discord and the Henderson-Vedral classical correlation are suitable measures
of correlations.

The correlations in the partition $AE_{A}$ are contained in the following
reduced-density operator, obtained by taking the partial trace over the
subsystems $B$ and $E_{B}$%
\end{subequations}
\begin{equation}
\rho_{AE_{A}}(p)=\frac{1}{2}%
\begin{bmatrix}
1 & 0 & 0 & 0\\
0 & 0 & 0 & 0\\
0 & 0 & 1-p & \sqrt{pq}\\
0 & 0 & \sqrt{pq} & p
\end{bmatrix}
,\label{r2d}%
\end{equation}
while for the partition $AE_{B}$ one obtains\begin{widetext}
\begin{equation}
\rho_{AE_{B}}\left(  p\right)  =\frac{1}{4}%
\begin{bmatrix}
1+q+pc_{3} & \left(  1-c_{3}\right)  \sqrt{pq} & 0 & 0\\
\left(  1-c_{3}\right)  \sqrt{pq} & \left(  1-c_{3}\right)  p & 0 & 0\\
0 & 0 & 1+q-pc_{3} & \left(  1+c_{3}\right)  \sqrt{pq}\\
0 & 0 & \left(  1+c_{3}\right)  \sqrt{pq} & \left(  1+c_{3}\right)  p
\end{bmatrix}.
\label{r3d}%
\end{equation}
\end{widetext}The last partition we want to analyze here is $E_{A}E_{B}$,
whose reduced-density operator of wich is given by \begin{widetext}
\begin{equation}
\rho_{E_{a}E_{b}}\left(  p\right)  =\frac{1}{4}%
\begin{bmatrix}
4q+\left(  1+c_{3}\right)  p^{2} & \gamma\sqrt{pq} & \gamma\sqrt{pq} & \left(
1+c_{3}\right)  pq\\
\gamma\sqrt{pq} & \gamma p & \left(  1+c_{3}\right)  pq & \left(
1+c_{3}\right)  p\sqrt{pq}\\
\gamma\sqrt{pq} & \left(  1+c_{3}\right)  pq & \gamma p & \left(
1+c_{3}\right)  p\sqrt{pq}\\
\left(  1+c_{3}\right)  pq & \left(  1+c_{3}\right)  p\sqrt{pq} & \left(
1+c_{3}\right)  p\sqrt{pq} & \left(  1+c_{3}\right)  p^{2}%
\end{bmatrix}
, \label{r4d}%
\end{equation}
\end{widetext}where $\gamma=2-\left(  1+c_{3}\right)  p$. For these last
states we must use the \textquotedblleft two-side\textquotedblright\ measures
of correlations (\ref{twosideCC}) and (\ref{twosideQC}). Before proceeding
with the numerical analysis, let us look at the entanglement between the
various partitions. Defining $\rho_{kl}^{T_{k}}$ as the partial transposition
of matrix $\rho_{kl}$ with respect to the subsystem $k$ \cite{Peres1}, we can
directly see that $\rho_{AE_{A}}^{T_{A}}\left(  p\right)  =\rho_{AE_{A}%
}\left(  p\right)  $, $\rho_{AE_{B}}^{T_{A}}\left(  p\right)  =\rho_{AE_{B}%
}\left(  p\right)  $, and $\rho_{E_{a}E_{b}}^{T_{E_{A}}}\left(  p\right)
=\rho_{E_{a}E_{b}}\left(  p\right)  $. From the Peres separability criterion
\cite{Peres1}, we see that there is no entanglement between the subsystems $A$
and $E_{A(B)}$ nor between the reservoirs $E_{A}$ and $E_{B}$, for any value
of the parametrized time $p$. Although no bipartite entanglement was observed
beyond that contained in the two-qubit initial state, multipartite
entanglement is always possible. To investigate this possibility, we consider
the following system-reservoir initial state:%
\[
\left\vert \psi_{in}\right\rangle =\frac{1}{\sqrt{2}}\left(  \left\vert
0\right\rangle _{A}\left\vert 1\right\rangle _{B}-\left\vert 1\right\rangle
_{A}\left\vert 0\right\rangle _{B}\right)  \otimes\left\vert 0\right\rangle
_{E_{A}}\left\vert 0\right\rangle _{E_{B}},
\]
which is obtained from Eq. (\ref{init}) by setting $c_{1}=c_{2}=c_{3}=-1$
(Werner state with $\alpha=1$). The action of the phase-damping channel
(\ref{Map2}) on the above state results in the following asymptotic ($p=1$)
system-reservoir state:%
\begin{equation}
\left\vert \psi_{a}\right\rangle =\frac{1}{\sqrt{2}}\left(  \left\vert
0\right\rangle _{A}\left\vert 1\right\rangle _{B}\left\vert 0\right\rangle
_{E_{A}}\left\vert 1\right\rangle _{E_{B}}-\left\vert 1\right\rangle
_{A}\left\vert 0\right\rangle _{B}\left\vert 1\right\rangle _{E_{A}}\left\vert
0\right\rangle _{E_{B}}\right)  .\label{astate}%
\end{equation}

Therefore, in the asymptotic limit, the system-reservoir state is a
quadripartite entangled state, the Greenberg-Horne-Zeilinger (GHZ) state
\cite{GHZ}. The GHZ class of states is the only one that possesses irreducible
multipartite correlations \cite{irreducible}. This means that the correlations
in state (\ref{astate}) cannot be determined by looking at its reduced-density
operators \cite{irreducible}. Retrieving to entanglement, there is no
tripartite or bipartite entanglements in state (\ref{astate}) because the
reduced-density operators are separable. This example indicates that although
the bipartite entanglement between $AE_{A}$ and $BE_{B}$ is null, multipartite
entanglement between all parts of the global system can be generated during
the decoherence process.

The next example illustrates another important feature of the two-qubit
dynamics under a phase-damping channel described previously. Consider a
separable system-reservoir initial state, for example $\rho_{ABE_{A}E_{B}%
}(0)=\rho_{AB}(0)\otimes\left\vert 0\right\rangle _{E_{A}}\left\langle
0\right\vert \otimes\left\vert 0\right\rangle _{E_{B}}\left\langle
0\right\vert $, $\rho_{AB}(0)$ being the Werner state with $\alpha\leq1/3$. As
the initial state is separable (and the environments are independent) and
there is interaction only between the partitions $AE_{A}$ and $BE_{B}$, no
multipartite entanglement is generated for all values of $p$. Therefore,
entanglement can only be created between the qubit $A$($B$) and its reservoir
$E_{A}$($E_{B}$) due to their interaction. However, as shown in Fig.6, the
qubit $A$($B$) does not get entangled with its reservoir for any value of $p$.
Although no bipartite or multipartite entanglement is created during the time
evolution, we do have decoherence, as can be seen from the asymptotic limit of
Eq. (\ref{r1d}). A possible explanation for this fact is the presence of
nonclassical correlations between the qubit and its reservoir (see Fig.6).
Decoherence without entanglement between the system and the reservoir were
noted earlier in the context of continuous variables \cite{Plenio}. On the
other hand, when one considers a single qubit under phase damping, the qubit
decoherence process is always accompanied by entanglement between the qubit
and its reservoir. This result was verified numerically for many qubit initial
states with nonzero coherence.

As we can see in Fig. 7, under phase damping, the quantum correlation
$\mathcal{Q}$ is asymptotically null, but the classical correlation
$\mathcal{K}$ reaches its maximum at this limit. Comparing this figure with
Fig. 6 , where the correlations for the partition $AE_{A}$ are plotted, we see
that the reduction of correlations in partition $AB$ is accompanied by the
creation of correlations in partition $AE_{A}$. We note that $\mathcal{Q}$
between the partitions $A(B)$ and $E_{A}(E_{B})$ starts to increase until it
reaches a maximum, decreasing to zero thereafter and leading, in the
asymptotic limit, to a classical correlated state between the qubits and its
reservoirs (the symmetry of the initial state leads to the same evolution for
the partition $BE_{B}$). In Fig. 8 we show the correlation dynamics for the
partition $AE_{B}$ where the same behavior as that of partition $AE_{A}$ is exhibited.

From Figs. 6 to 8, we observe that the quantum correlations (including the
initial entanglement) disappear in the asymptotic regime ($p=1$), in all
partitions considered. This is not the case for the amplitude-damping channel,
where the quantum correlations are completely transferred from the system $AB$
to the reservoirs $E_{A}E_{B}$ at $p=1$.

%

\begin{figure}
[h]
\begin{center}
\includegraphics[
natheight=8.437100in,
natwidth=10.562800in,
height=7.2862cm,
width=9.1006cm
]%
{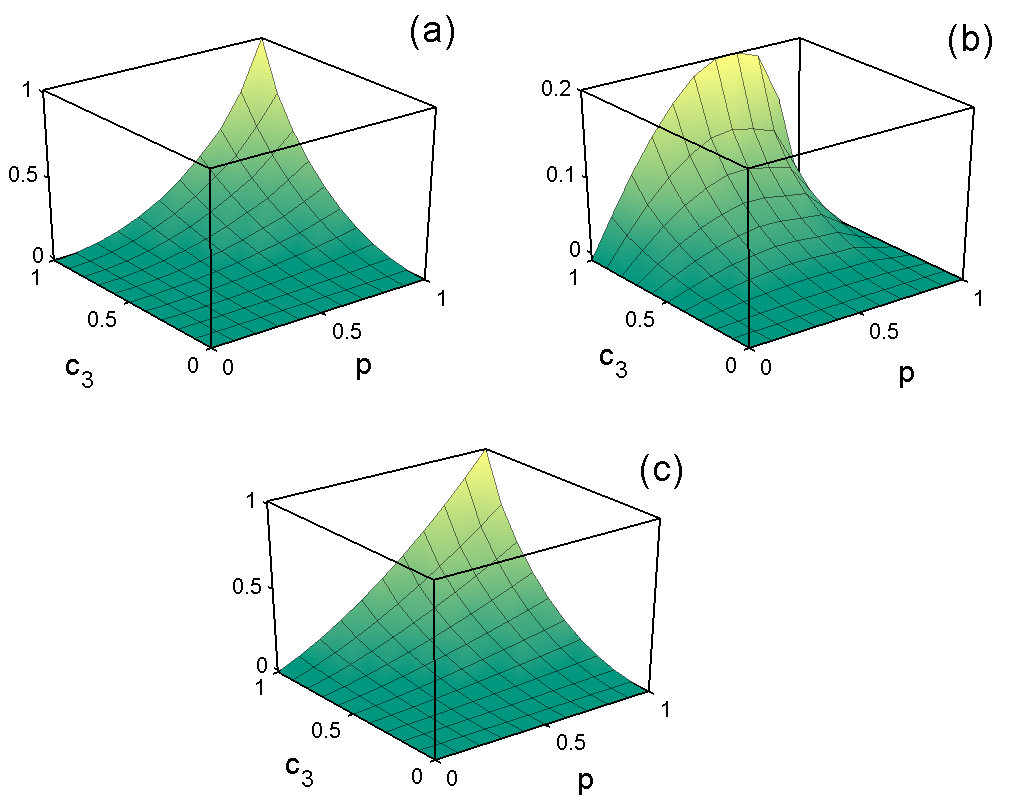}%
\caption{(Color online) Correlation dynamics for the dephasing channel
considering the partition $AE_{A}$, for the general state (\ref{init}). (a)
Classical correlation [Eq. (\ref{twosideCC})], (b) quantum correlation [Eq.
(\ref{twosideQC})] and (c) mutual information [Eq. (\ref{QMI1})].}%
\end{center}
\end{figure}
%

\begin{figure}
[h]
\begin{center}
\includegraphics[
natheight=10.666600in,
natwidth=13.760000in,
height=5.7398cm,
width=8.0946cm
]%
{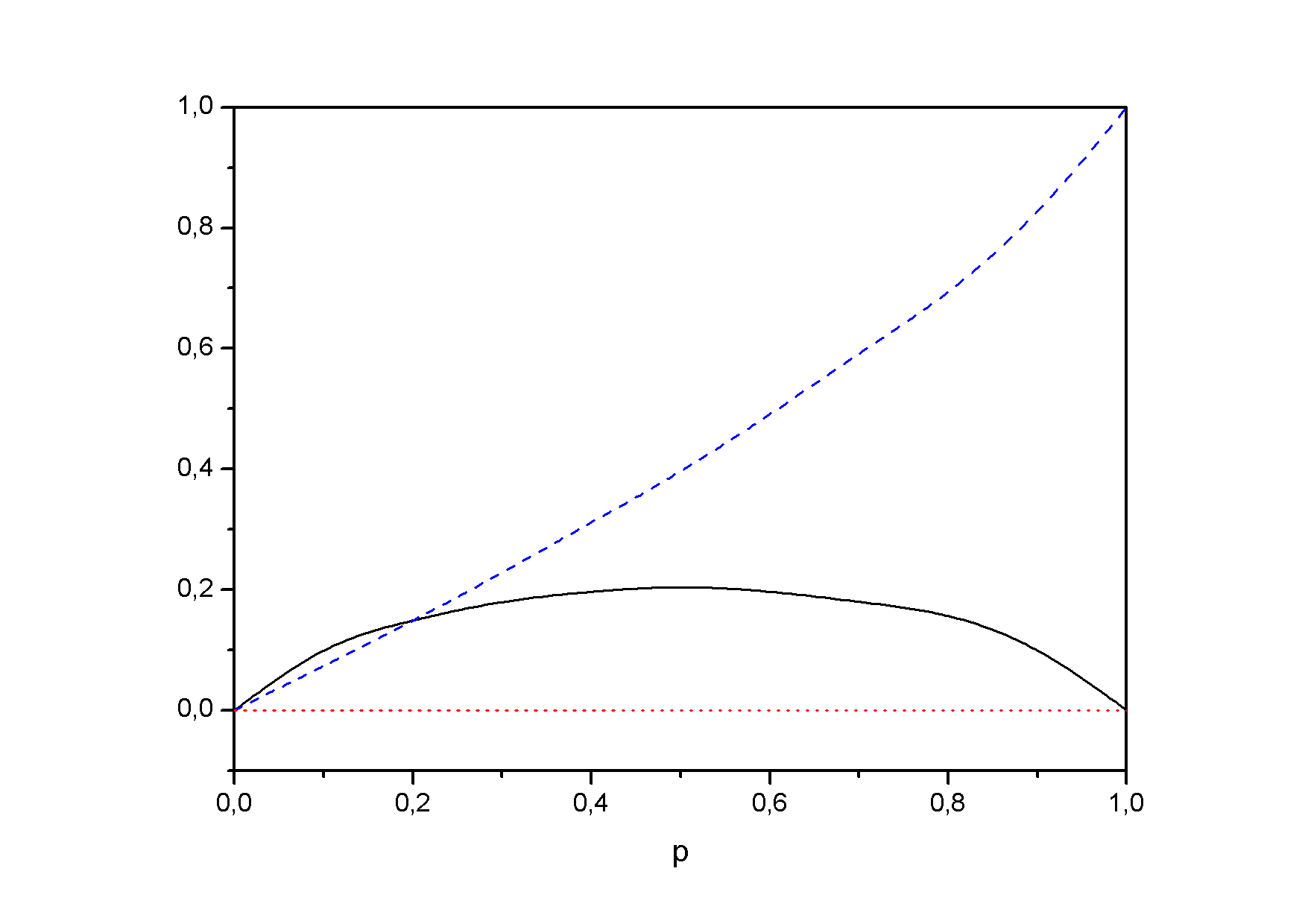}%
\caption{(Color online) Correlation dynamics for the dephasing channel,
considering the partition $AE_{A}$, for the general state (\ref{init}).
Classical correlation (dashed line) given by Eq. (\ref{twosideCC}), quantum
correlation (solid line) given by Eq. (\ref{twosideQC}) and concurrence
(dotted line) given by Eq. (\ref{Conc1}).}%
\end{center}
\end{figure}
%

\begin{figure}
[h]
\begin{center}
\includegraphics[
natheight=8.312600in,
natwidth=10.437400in,
height=7.2642cm,
width=9.1028cm
]%
{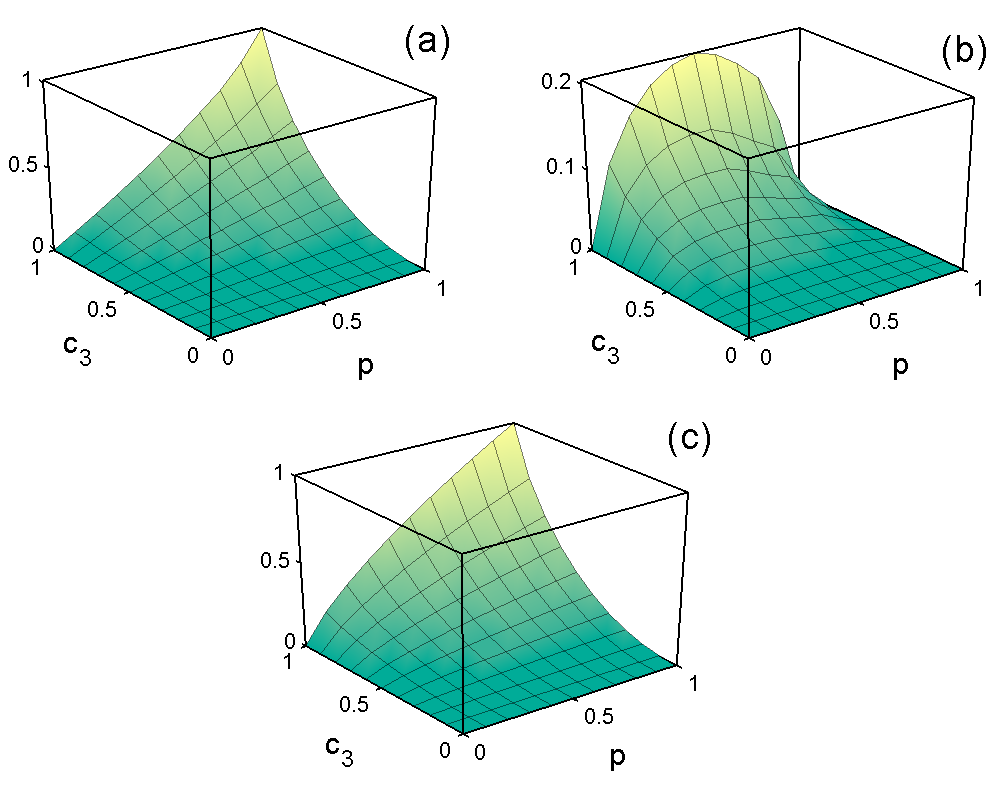}%
\caption{(Color online) Correlation dynamics for the dephasing channel,
considering partition $AE_{B}$, for the general state (\ref{init}). (a)
classical correlation [Eq. (\ref{twosideCC})], (b) quantum correlation [Eq.
(\ref{twosideQC})] and (c) mutual information [Eq. (\ref{QMI1})].}%
\end{center}
\end{figure}

\subsection{Bit-Flip, bit-phase-flip and phase-flip channels}

The effect of bit-flip, bit-phase-flip and phase-flip channels is to destroy
the information contained in the phase relations without an exchange of
energy. The action of these channels on a single qubit can be described by the
following Kraus operators%
\begin{equation}
\Gamma_{0}^{(k)}=\sqrt{q^{\prime}}%
\begin{bmatrix}
1 & 0\\
0 & 1
\end{bmatrix}
\text{, \ }\Gamma_{1}^{(k)}=\sqrt{p}\sigma_{i}^{(k)}\text{,}\label{KBPBPF}%
\end{equation}
where $i=1$ ($x$ axis) for the bit-flip, $i=2$ ($y$ axis) for the
bit-phase-flip, $i=3$ ($z$ axis) for the phase flip ($k=A,B$), and we defined
$q^{\prime}=1-p/2$.

It is helpful to get a geometrical picture by looking at the Bloch sphere
representation of one qubit \cite{NieChu}. To this end, let us then consider
the action of the bit-flip channel. Owing to the symmetry of the Kraus
operator (it is proportional to $\sigma_{x}$), all the points on the sphere
are uniformly compressed on the $x$ axis. Then states on this axis will be
invariant under the bit-flip channel, as can be seen directly from Eq.
(\ref{KBPBPF}). It is not difficult to see that the actions of the other two
channels are completely equivalent to that of the bit flip, the only
difference being the symmetry axis. The bit-phase flip channel will leave
invariant states on the $y$ axis, while for the phase flip channel, the
symmetry axis is the $z$ axis. For this reason, we will present here the case
of the bit flip channel alone. For completeness, the phase-flip and
bit-phase-flip channels are presented in the Appendix A.

Considering the initial state (\ref{init}), the evolved reduced density matrix
for the partition $AB$ under bit flip is given by%
\begin{equation}
\varepsilon(\rho_{AB})=\frac{1}{4}%
\begin{bmatrix}
1+c_{3}q^{2} & 0 & 0 & c_{1}-c_{2}q^{2}\\
0 & 1-c_{3}q^{2} & c_{1}+c_{2}q^{2} & 0\\
0 & c_{1}+c_{2}q^{2} & 1-c_{3}q^{2} & 0\\
c_{1}-c_{2}q^{2} & 0 & 0 & 1+c_{3}q^{2}%
\end{bmatrix}
. \label{r1bf}%
\end{equation}

Once more, we can use \textquotedblleft one-side\textquotedblright\ measures
of classical (\ref{HV}) and quantum (\ref{Dis}) correlations, which can be
computed analytically for this case \cite{Maziero}. They are given by Eqs.
(\ref{CC}) and (\ref{CQ}), but with $\chi=\max\left\{  \left\vert
c_{1}\right\vert ,q^{2}\left\vert c_{2}\right\vert ,q^{2}\left\vert
c_{3}\right\vert \right\}  $. Note that the axes $y$ and $z$ are continuously
contracted by the factor $q^{2}$ while the $x$ axes is left invariant.

The bipartitions of subsystem $A$ and each of the two reservoirs are given by%
\begin{equation}
\rho_{AE_{A}}(p)=\frac{1}{2}%
\begin{bmatrix}
q^{\prime} & 0 & 0 & \sqrt{pq^{\prime}/2}\\
0 & p/2 & \sqrt{pq^{\prime}/2} & 0\\
0 & \sqrt{pq^{\prime}/2} & q^{\prime} & 0\\
\sqrt{pq^{\prime}/2} & 0 & 0 & p/2
\end{bmatrix}
,\label{r2bf}%
\end{equation}
and%
\begin{equation}
\rho_{AE_{B}}\left(  p\right)  =\frac{1}{2}%
\begin{bmatrix}
q^{\prime} & 0 & 0 & c_{1}\sqrt{pq^{\prime}/2}\\
0 & p/2 & c_{1}\sqrt{pq^{\prime}/2} & 0\\
0 & c_{1}\sqrt{pq^{\prime}/2} & q^{\prime} & 0\\
c_{1}\sqrt{pq^{\prime}/2} & 0 & 0 & p/2
\end{bmatrix}
.\label{r3bf}%
\end{equation}
From these equations we directly see that $\rho_{AE_{A}}^{T_{A}}\left(
p\right)  =\rho_{AE_{A}}\left(  p\right)  $ and $\rho_{AE_{B}}^{T_{A}}\left(
p\right)  =\rho_{AE_{B}}\left(  p\right)  $ implying, once more, from the
Peres separability criteria \cite{Peres1} that we have decoherence without
entanglement between the qubits and the reservoirs for any parametrized time
$p$. The last partition is given by%
\begin{equation}
\rho_{E_{A}E_{B}}\left(  p\right)  =%
\begin{bmatrix}
\left(  q^{\prime}\right)  ^{2} & 0 & 0 & c_{1}pq^{\prime}/2\\
0 & pq^{\prime}/2 & c_{1}pq^{\prime}/2 & 0\\
0 & c_{1}pq^{\prime}/2 & pq^{\prime}2 & 0\\
c_{1}pq^{\prime}/2 & 0 & 0 & p^{2}/4
\end{bmatrix}
,\label{r4bf}%
\end{equation}
which also has the property $\rho_{E_{A}E_{B}}^{T_{E_{A}}}\left(  p\right)
=\rho_{E_{A}E_{B}}\left(  p\right)  $. Figure 9 shows similar behavior to that
in Fig. 6 for the phase-damping channel. During the process of decoherence,
subsystem $A$ becomes quantum correlated (but not entangled) with its own
reservoir and the asymptotic state possesses only classical correlations.%

\begin{figure}
[h]
\begin{center}
\includegraphics[
natheight=10.666600in,
natwidth=13.760000in,
height=5.7398cm,
width=8.0946cm
]%
{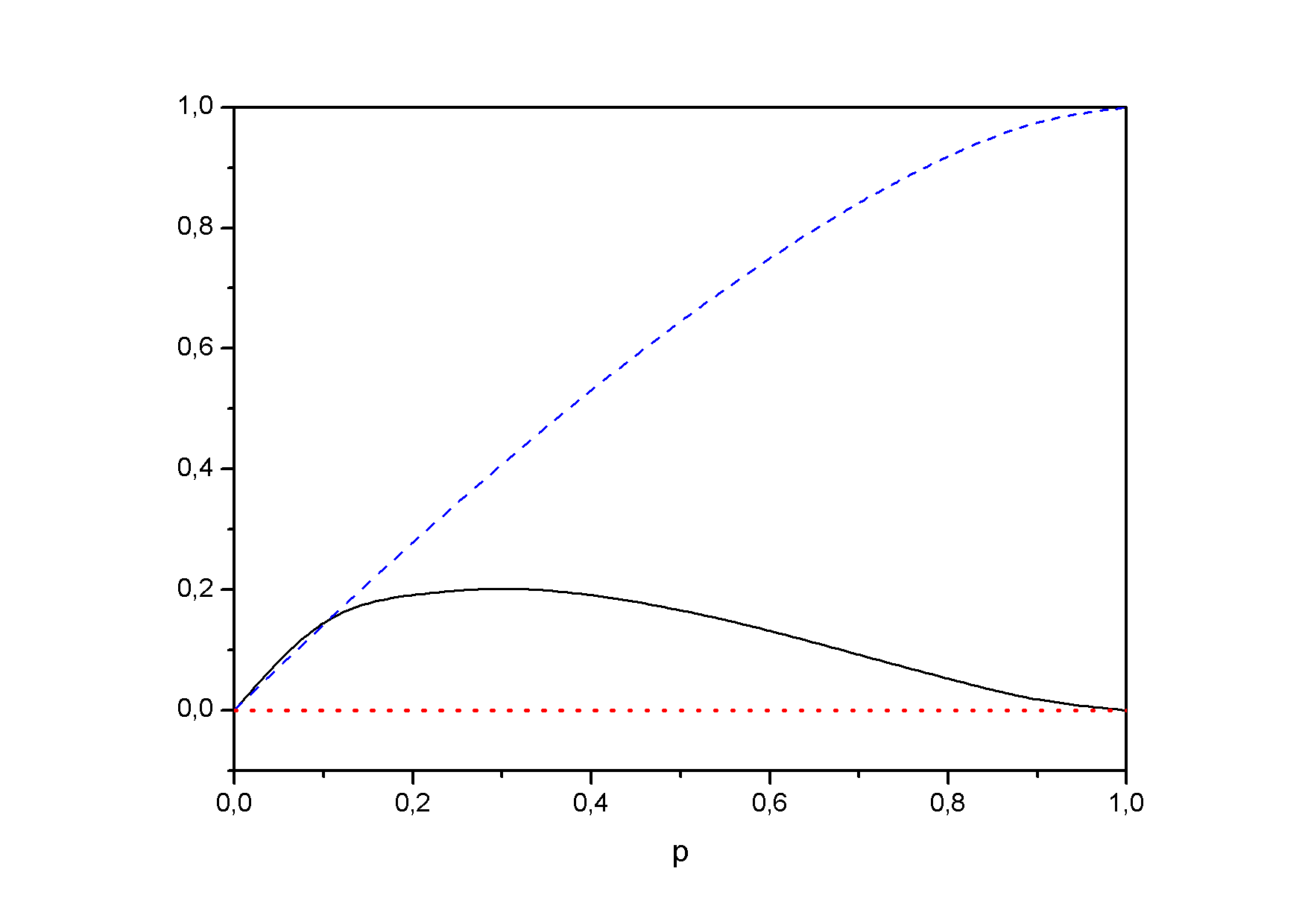}%
\caption{(Color online) Correlation dynamics for the bit flip channel,
considering the partition $AE_{A}$, for the general state (\ref{init}).
Classical correlation (dashed line) given by Eq. (\ref{twosideCC}), quantum
correlation (solid line) given by Eq. (\ref{twosideQC}) and the concurrence
(dotted line) given by Eq. (\ref{Conc1}).}%
\end{center}
\end{figure}

For completeness, in Figs. 10 and 11 we plot the correlation dynamics for the
partitions $AE_{B}$ and $E_{A}E_{B}$, respectively. As we can see, the dynamic
behavior of the correlations under the bit-flip channel is essentially the
same as that under the phase-damping channel (Figs. 6 and 8).%

\begin{figure}
[ptb]
\begin{center}
\includegraphics[
natheight=8.312600in,
natwidth=10.437400in,
height=7.2642cm,
width=9.1028cm
]%
{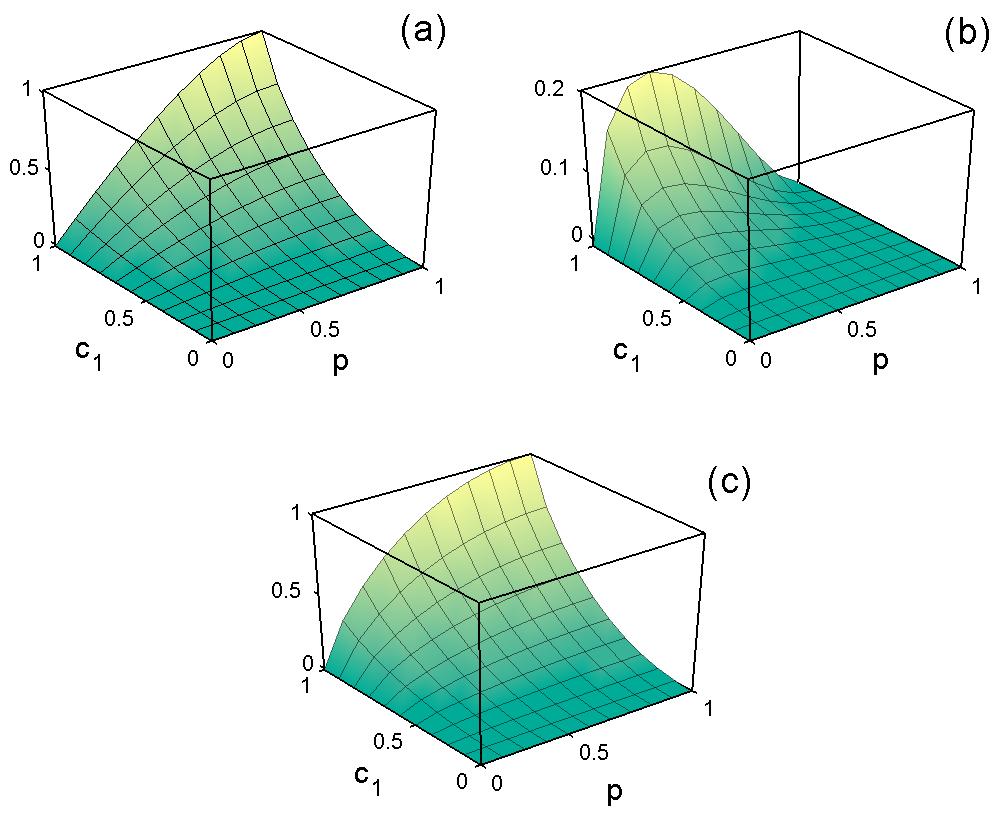}%
\caption{(Color online) Correlation dynamics for the bit flip channel,
considering partition $AE_{B}$, for the general state (\ref{init}). (a)
Classical correlation [Eq. (\ref{twosideCC})], (b) quantum correlation [Eq.
(\ref{twosideQC})], and (c) mutual information [Eq. (\ref{QMI1})].}%
\end{center}
\end{figure}
%

\begin{figure}
[ptb]
\begin{center}
\includegraphics[
natheight=8.437100in,
natwidth=10.562800in,
height=7.2862cm,
width=9.1006cm
]%
{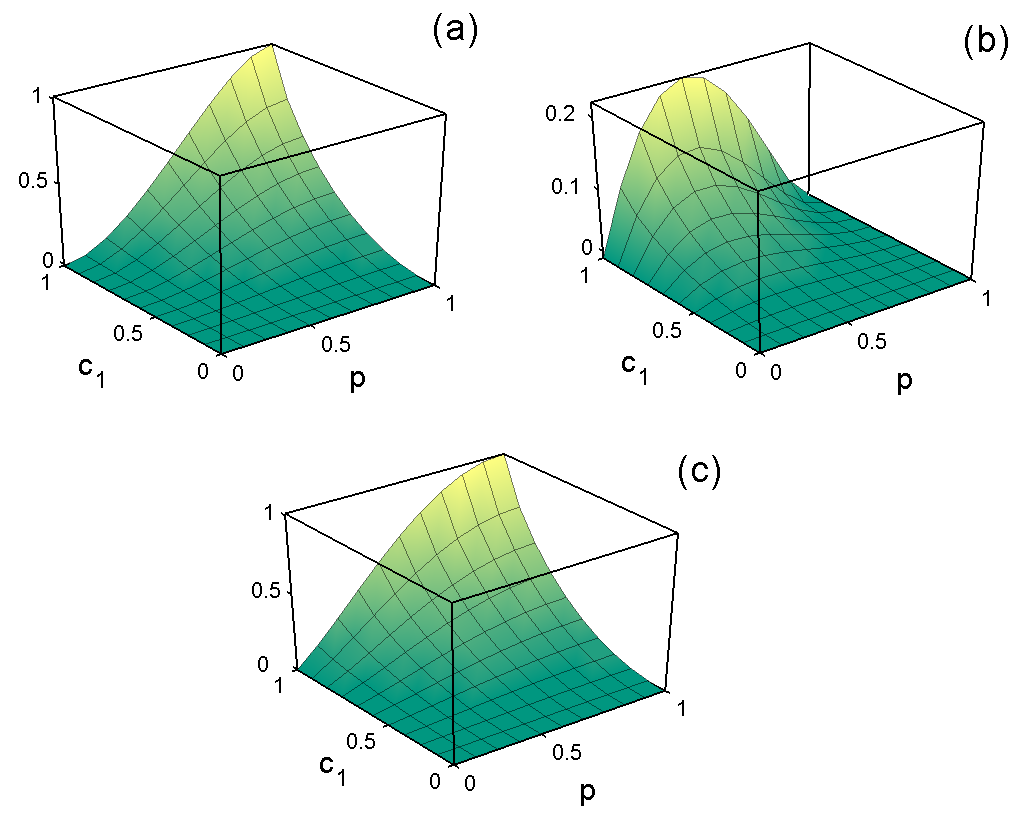}%
\caption{(Color online) Correlation dynamics for the bit flip channel,
considering partition $E_{A}E_{B}$, for the general state (\ref{init}). (a)
classical correlation [Eq. (\ref{twosideCC})], (b) quantum correlation [Eq.
(\ref{twosideQC})], and (c) mutual information [Eq. \ref{QMI1})].}%
\end{center}
\end{figure}

\section{Summary and concluding remarks}

In this article we investigate the system-reservoir dynamics of both classical
and quantum correlations in the decoherence phenomenon. We considered all
possible bipartitions of a two-qubit system interacting with two local,
independent environments, modeling several common noise sources:
amplitude-damping, phase-damping, bit-flip, bit-phase-flip, and phase-flip channels.

We observe here two distinct behaviors for the dynamics of correlations, when
the qubits are under the action of (i) amplitude-damping and (ii)
phase-damping, bit-flip, bit-phase-flip and phase-flip channels. In case (i),
all correlations (classical and quantum, including entanglement) initially
present in the system are completely transferred in the asymptotic time, to
the environments. During time evolution, all bipartitions of the complete
system exhibit some degree of correlation, including entanglement. In case
(ii), the classical and quantum correlations initially present in the system
are transferred over time to all bipartitions of the complete system, but the
entanglement is not transferred. The bipartite entanglement contained in the
system is completely \textquotedblleft evaporated\textquotedblright\ by the
action of the channels. While all bipartitions develop quantum correlation of
separable states during evolution, at the asymptotic time all nonclassical
bipartite correlations are null. Thus, the asymptotic state of the whole
system (system of interest plus environment) contains only classical
correlation in all bipartitions. In case (ii), we have decoherence without
entanglement between the qubits and the environment, the classical and quantum
correlations (of separable states) are responsible for the information
transfer from the system to the environment.

Finally we note that we study here only bipartite correlations. Certainly a
study of multipartite correlations will make a useful contribution to
understanding the dynamics of information in the decoherence process. An
important future investigation will be the study of the effects of
finite-temperature environments on the dynamics of these correlations. Another
interesting line of research is the dynamic behavior of the system under the
action of a single environment, where correlations can be created in the
system through nonlocal interactions mediated by the environment.

\begin{acknowledgments}
We are grateful for the funding from UFABC, CAPES, FAPESP, CNPq, and the
Brazilian National Institute for Science and Technology of Quantum Information (INCT-IQ).
\end{acknowledgments}

\appendix{}

\section{Correlation Dynamics for bit-phase-flip and phase-flip Channels}

The $AB$ reduced-density matrix for the bit-phase-flip channel is given by%
\[
\rho_{AB}\left(  p\right)  =\frac{1}{4}%
\begin{bmatrix}
1+c_{3}q^{2} & 0 & 0 & c_{1}q^{2}-c_{2}\\
0 & 1-c_{3}q^{2} & c_{1}q^{2}+c_{2} & 0\\
0 & c_{1}q^{2}+c_{2} & 1-c_{3}q^{2} & 0\\
c_{1}q^{2}-c_{2} & 0 & 0 & 1+c_{3}q^{2}%
\end{bmatrix}
,
\]
the classical and quantum correlations being given by Eqs. (\ref{CC}) and
(\ref{CQ}), with $\chi=\max\left\{  q^{2}\left\vert c_{1}\right\vert
,\left\vert c_{2}\right\vert ,q^{2}\left\vert c_{3}\right\vert \right\}  $. We
see that the entire Bloch sphere is shrunk onto the $y$ axis, which is the
symmetry axis for the bit-phase-flip channel. The density operator for the
partition $AE_{A}$ is given by%
\[
\rho_{AE_{A}}\left(  p\right)  =\frac{1}{2}%
\begin{bmatrix}
q^{\prime} & 0 & 0 & -i\sqrt{pq^{\prime}/2}\\
0 & p/2 & -i\sqrt{pq^{\prime}/2} & 0\\
0 & i\sqrt{pq^{\prime}/2} & q^{\prime} & 0\\
i\sqrt{pq^{\prime}/2} & 0 & 0 & p/2
\end{bmatrix}
\]
and for the partition $AE_{B}$, by \begin{widetext}
\[
\rho_{AE_{B}}\left(  p\right)  =\frac{1}{2}%
\begin{bmatrix}
1-p/2 & 0 & 0 & -ic_{2}\sqrt{pq^{\prime}/2}\\
0 & p/2 & -ic_{2}\sqrt{pq^{\prime}/2} & 0\\
0 & ic_{2}\sqrt{pq^{\prime}/2} & q^{\prime} & 0\\
ic_{2}\sqrt{pq^{\prime}/2} & 0 & 0 & p/2
\end{bmatrix}
,
\]
\end{widetext}where we directly see from the Peres partial transposition
criterion \cite{Peres1} that there is no bipartite entanglement in these
cases. The last bipartition reads%
\[
\rho_{E_{a}E_{b}}\left(  p\right)  =%
\begin{bmatrix}
\left(  q^{\prime}\right)  ^{2} & 0 & 0 & c_{2}pq^{\prime}/2\\
0 & pq^{\prime}/2 & c_{2}pq^{\prime}/2 & 0\\
0 & c_{2}pq^{\prime}/2 & pq^{\prime}2 & 0\\
c_{2}pq^{\prime}/2 & 0 & 0 & p^{2}/4
\end{bmatrix}
.
\]
From this last equation we see once more that the reservoir never gets entangled.

For the phase-flip channel, the partition $AB$ is%
\[
\rho_{AB}\left(  p\right)  =\frac{1}{4}%
\begin{bmatrix}
1+c_{3} & 0 & 0 & c^{-}q^{2}\\
0 & 1-c_{3} & c^{+}q^{2} & 0\\
0 & c^{+}q^{2} & 1-c_{3} & 0\\
c^{-}q^{2} & 0 & 0 & 1+c_{3}%
\end{bmatrix}
,
\]
the classical and quantum correlations being given by Eqs. (\ref{CC}) and
(\ref{CQ}), with $\chi=\max\left\{  q^{2}\left\vert c_{1}\right\vert
,q^{2}\left\vert c_{2}\right\vert ,\left\vert c_{3}\right\vert \right\}  $. We
see that the symmetry axis for the bit-phase-flip channel is the $z$ axis (as
for dephasing) --- the Bloch sphere is compressed onto this axis. The density
operator for the partition $AE_{A}$ is given by%
\[
\rho_{AE_{A}}\left(  p\right)  =\frac{1}{2}%
\begin{bmatrix}
1-p/2 & \sqrt{pq^{\prime}/2} & 0 & 0\\
\sqrt{pq^{\prime}/2} & p/2 & 0 & 0\\
0 & 0 & q^{\prime} & -\sqrt{pq^{\prime}/2}\\
0 & 0 & -\sqrt{pq^{\prime}/2} & p/2
\end{bmatrix}
\]
and for the partition $AE_{B}$ is \begin{widetext}
\[
\rho_{AE_{B}}\left(  p\right)  =\frac{1}{2}%
\begin{bmatrix}
1-p/2 & c_{3}\sqrt{pq^{\prime}/2} & 0 & 0\\
c_{3}\sqrt{pq^{\prime}/2} & p/2 & 0 & 0\\
0 & 0 & q^{\prime} & -c_{3}\sqrt{pq^{\prime}/2}\\
0 & 0 & -c_{3}\sqrt{pq^{\prime}/2} & p/2
\end{bmatrix}
,
\]
\end{widetext}where we directly see from the Peres partial transposition
criterion \cite{Peres1} that there is no bipartite entanglement in these
cases. The last bipartition reads%
\[
\rho_{E_{a}E_{b}}\left(  p\right)  =%
\begin{bmatrix}
\left(  q^{\prime}\right)  ^{2} & 0 & 0 & c_{3}pq^{\prime}/2\\
0 & pq^{\prime}/2 & c_{3}pq^{\prime}/2 & 0\\
0 & c_{3}pq^{\prime}/2 & pq^{\prime}2 & 0\\
c_{3}pq^{\prime}/2 & 0 & 0 & p^{2}/4
\end{bmatrix}
,
\]
which also does not exhibit entanglement.

\end{document}